# How a polymer filling enhances the rate and selectivity of colloid permeation across mesopores

Mikhail Y. Laktionov[1], Frans A. M. Leermakers[2], Ralf P. Richter[3*], Leonid I. Klushin[4,5*], Oleg V. Borisov[1*]

[1]CNRS, Université de Pau et des Pays de l'Adour, UMR 5254, Institut des Sciences Analytiques et de Physico-Chimie pour l'Environnement et les Matériaux, 64053 Pau, France.

[2]Physical Chemistry and Soft Matter, Wageningen University, Stippeneng 4, 6708 WE, Wageningen, The Netherlands.

[3]University of Leeds, School of Biomedical Sciences, Faculty of Biological Sciences; School of Physics and Astronomy, Faculty of Engineering and Physical Sciences; Astbury Centre for Structural Molecular Biology; and Bragg Centre for Materials Research, Leeds LS2 9JT, United Kingdom.

[4]Branch of Petersburg Nuclear Physics Institute named by B.P. Konstantinov of National Research Centre "Kurchatov Institute", Institute of Macromolecular Compounds, 199004 St. Petersburg, Russia.

[5]American University of Beirut, Department of Physics, Beirut 1107 2020, Lebanon.

*Corresponding authors: Ralf P. Richter (r.richter@leeds.ac.uk); Leonid I. Klushin (leo@aub.edu.lb); Oleg V. Borisov (oleg.borisov@univ-pau.fr)

**Polymer-functionalised mesopores are an emerging technology for colloid separation, sensing and delivery. Their potential is strikingly illustrated in living cells, where nuclear pore complexes (NPCs) control biocolloid transport between the nucleus and the cytosol. Even colloids much smaller than the biopolymer-filled NPC channel are effectively blocked, but some larger colloids with distinct surface features rapidly permeate. Simplistically, one may expect any polymer filling to obstruct and slow down colloid transport. We demonstrate how a polymer filling that attracts colloids and extends beyond the mesopore, thus maximizing colloid capture, can instead increase permeation compared to a bare pore. We also define how polymer-filled mesopores can effectively gate colloids according to their size and surface features. Our findings provide a basic physical explanation for the exquisite permselectivity of NPCs, and a rational design strategy for novel mesopore-based separation, sensing, catalysis and drug delivery devices with enhanced performance features.**

Advances in macromolecular chemistry have enabled the functionalization of mesopores (i.e., pores with a diameter between a few and many tens of nanometers) by anchoring polymer chains of various chemical nature to the pore wall. This creates a solvated polymer meshwork that fills the entire pore volume or remains confined to near-wall regions, depending on the polymer's molecular weight and conformational state (*1*, *2*). Polymer-modified mesoporous materials and membranes represent a new class of functional nanostructured systems with significant potential across a wide range of technologies ((*3*, *4*) and references therein), such as ultrafiltration, sensing, catalysis and drug delivery.

The interaction of the polymer meshwork with colloids, i.e., nanoparticles or (macro)molecules in the solution phase, essentially determines the absorption and separation properties of the polymer-modified mesoporous materials. Interactions can be attractive or repulsive, and of varying specificity, controlled by a broad spectrum of parameters (*5*) such as pH and ion strength, valency and specificity (*6*), solvent quality and temperature (*7*, *8*), and more specific 'ligand binding' features. This opens up novel opportunities for highly selective and controlled uptake and transport of colloids through polymer-filled mesoscopic channels (*9*, *10*).

Past experimental and theoretical efforts have focused on triggering the transient opening of a polymer free path through an external stimulus to gate transport (*10–13*). However, the polymer phase itself can potentially also provide high selectivity to colloids as a function of their size and attraction by the polymer. We thus hypothesized that even mesopores filled with a polymer meshwork across their entire cross-section can effectively gate transport. If successful, this approach would enable more robust gating as it does not rely on careful tuning of the diameter of a polymer-free channel, and higher transport rates as the full pore cross-section can participate in colloid transport.

Nature provides a case in point. Nuclear pore complexes (NPCs) perforate the nuclear envelope of eukaryotic cells and provide for speedy and highly selective nucleo-cytoplasmic transport of proteins and nucleic acids. This process spatially separates gene transcription (in the nucleus) from translation into proteins (in the cytosol), critical for the ordered course of gene expression. Each NPC forms an approximately cylindrical channel, measuring 40-80 nm in diameter (depending on cell state and species (*14*, *15*)) and similar in length. The channel is filled with a meshwork of hundreds of natively disordered protein domains rich in phenylalanine-glycine dipeptides (FG domains) that are anchored to the channel walls (*14*, *16–18*). Collectively, the FG domains provide remarkable gating function: biocolloids of just a tenth of the pore diameter or more in hydrodynamic diameter are effectively blocked, except for dedicated transport factors (importins and exportins, alone and in complex with cargo) that bind to the FG domains and permeate rapidly (*19*).

Several independent strands of evidence indicate that diffusive transport across NPCs is based on rather universal physical principles, whereas the exact chemical makeup of the polymers and colloids is secondary for function (*19*). First, NPCs robustly fulfill their transport functions despite substantial compositional variations of the NPC architecture across distant eukaryotic taxa and cell states (*14*, *16–18*, *20*, *21*). Second, NPCs can gate diffusive colloid transport similarly well in both directions. Whilst the native cell is capable of directed transport of cargo against a concentration gradient, this function is not intrinsic to the NPC and reversible through cell engineering (*22*). Third, the binding behaviour of transport factors to assemblies of purified FG domains could be reproduced by simple physical models that treat FG domains as regular flexible polymers and transport factors as spherical colloids with a homogeneous surface (*23*, *24*), i.e., ignoring the detailed arrangement of interaction sites. Fourth, studies with a spectrum of globular proteins as model colloids demonstrated that NPCs exhibit a wide and continuous spectrum of permeabilities as a function of protein surface properties (at constant size (*25*)) and size (for weakly or non-interacting proteins (*26*–*29*)). Based on these and other findings, it is thought that a fine balance of many individually weak physicochemical (e.g., electrostatic, hydrophobic, aromatic stacking) interactions between polymers and biocolloids dictates the gating behaviour, rather than a few highly specific biochemical interactions (*19*).

However, we currently lack an understanding of the relationship between the molecular architecture of the polymer brush filling the pore, and its ability to transport colloids with high selectivity and rate. Here, we address this question with a self-consistent field theoretical approach. The theory considers that a meshwork of flexible polymers effectively increases the local viscosity and thereby slows down transport of colloids compared to an open pore. On the other hand, an attractive polymer phase recruits colloids into the pore, thus increasing colloid transport. Intriguingly, the solvent strength through its influence on the density and compactness of the polymer meshwork impacts both these effects. We define how solvent quality and colloid attraction to the polymer may be tuned to maximize the transport rate (even beyond the rate for an open pore) and to achieve highly selective transport with respect to the colloid's size or affinity for the polymer.

## Results

### Defining the transport scenario

Salient features of our simulated mesopore are illustrated in Figure 1. The cylinder-shaped pore perforates a planar membrane (for the NPC this would be space bounded by two lipid bilayers) and is the sole conduit for colloids between two semi-infinite solution reservoirs. Flexible polymer chains are end-grafted to the inner pore walls, at a density sufficient to form a polymer brush that fills the entire pore cross-section.

We will focus on a pore with a set radius $r_\mathrm{p}^0$ and length $L_0$ (in units $a$), and polymers with a degree of polymerization $N$ and grafting density $\sigma$ (Figure 1). Whilst the selected values are inspired by the NPC (see Supplementary Note 1), we expect that our findings will be of rather general validity so they can be applied to the performance analysis and rational design of mesopores with other geometries or polymer fillings.

Colloids are taken to be spheres with diameter $d$. The interaction strength (contact free energy)

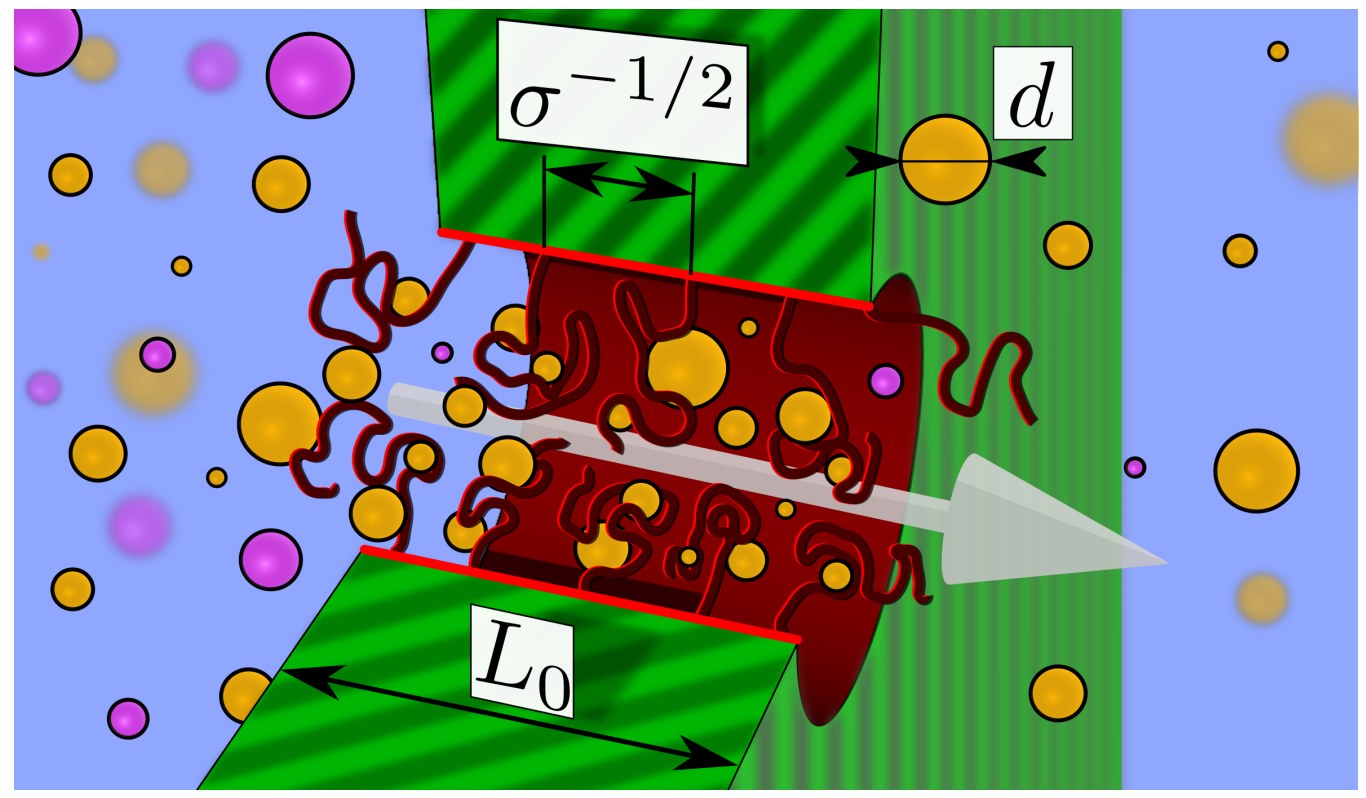

**Figure 1**: **Schematic of colloid diffusive transport through a pore filled with a polymer brush.** The brush is formed by linear polymer chains (red strands) with a degree of polymerization $N$, uniformly grafted with grafting density $\sigma$ to the inner surface of a cylindrical pore. The pore radius is $r_\mathrm{p}^0$ (not indicated) and the thickness of the impermeable membrane is $L_0$. Polymer chains are flexible with a statistical segment length $a$ and volume $\sim a^3$. Spherical colloids with diameter $d$ are free to diffuse in the surrounding solvent. All length scales are normalized by $a$. As a model pore, we set $L_0 = 2r_\mathrm{p}^0 = 56$, $\sigma = 0.02$ and $N = 300$. With $a = 0.76$ nm, these parameters approximate basic features of a NPC.

between a polymer segment and the colloid surface is represented by the Flory-Huggins parameter $\chi_\mathrm{PC}$.

To understand how the polymer brush affects transport, we consider the stationary diffusive flux of colloids through the pore and analyze how it depends on the parameters of the pore, the brush, and the colloid. We consider unidirectional colloid transport driven solely by the concentration difference across the membrane and focus on the fundamental mechanisms of diffusion mediated by colloid-polymer interactions. Far away from the membrane, the colloid concentrations are set to $c = c_0$ (at $z \to -\infty$) and $c = 0$ (at $z \to +\infty$).

### Stationary flux

Diffusive transport in the presence of the external mean force generated by the position-dependent insertion free energy $\Delta F(\boldsymbol{r})$ (in units of thermal energy $k_\mathrm{B}T$) is described by the Smoluchowski equation,

$$\frac{\partial c(\boldsymbol{r})}{\partial t} = -\nabla \cdot \left[D(\boldsymbol{r})\nabla c(\boldsymbol{r}) + D(\boldsymbol{r})c(\boldsymbol{r})\nabla\big(\Delta F(\boldsymbol{r})\big)\right], \tag{1}$$

where $c(\boldsymbol{r})$ is the concentration of the colloid, and $D(\boldsymbol{r})$ is its position-dependent diffusion coefficient. Standard substitution introduces the potential function $\psi(\boldsymbol{r})$ and the effective conductivity $\tilde{D}(\boldsymbol{r})$ as

$$\psi(\boldsymbol{r}) = c(\boldsymbol{r}) \exp(\Delta F(\boldsymbol{r})) \tag{2}$$

$$\tilde{D}(\boldsymbol{r}) = D(\boldsymbol{r}) \exp(-\Delta F(\boldsymbol{r})) \tag{3}$$

With these substitutions, the flux density is expressed as

$$\boldsymbol{j} = -\tilde{D}(\boldsymbol{r})\nabla\psi(\boldsymbol{r}) \tag{4}$$

Under stationary conditions, Eq. (1) is reduced to

$$\nabla \cdot \big(\tilde{D}(\boldsymbol{r})\nabla\psi(\boldsymbol{r})\big) = 0 \tag{5}$$

The boundary conditions for the potential function in our case are $\psi(z \to -\infty) = c_0$ and $\psi(z \to +\infty) = 0$ since the insertion free energy $\Delta F$ vanishes far away from the pore. The normal component of the flux density vanishes at the impenetrable membrane walls. Once the insertion free energy field $\Delta F(\boldsymbol{r})$ and the position-dependent diffusion coefficient $D(\boldsymbol{r})$ are defined, a natural approach to find the total flux (resistance) is by direct numerical solution of Eqs. (3-5) which we did as a check (see below). However, a more transparent and instructive approach is to follow an electric analog of the problem which can be reformulated as finding the total resistance of a medium with position-dependent resistivity possessing axial symmetry:

$$\rho(r, z) = \tilde{D}^{-1}(r, z) \tag{6}$$

The variation of the local resistivity $\rho(r, z)$ is due to the position-dependent polymer volume fraction $\phi(r, z)$

**Bare pore as a reference case.** A natural reference is the diffusive flux through a bare pore, which itself limits the transport of solutes (*30*, *31*). The insertion free energy is $\Delta F = 0$ everywhere, so that the potential function $\psi(\boldsymbol{r})$ coincides with $c(\boldsymbol{r})$, and the diffusion coefficient is position independent.

Lord Rayleigh analyzed the flux of solute particles of negligible size through a circular pore in a planar membrane of negligible thickness (*32*). In this simplest case, the equiconcentration surfaces are oblate spheroids, and the streamlines form confocal hyperboloids of revolution (*33*). The net flux through the pore is

$$J = 2D_0 r_{\mathrm{p}} \Delta c, \tag{7}$$

where $r_{\mathrm{p}}$ is the pore radius and $D_0 = k_{\mathrm{B}}T/(3\pi\eta_{\mathrm{S}}d)$ is the diffusion coefficient of the colloid in plain solvent with viscosity $\eta_{\mathrm{S}}$. Negligible colloid size means $d \ll r_{\mathrm{p}}$.

A membrane of finite thickness $L$ allows the approximate analytical solution (*34*)

$$J = \frac{2D_0 r_{\mathrm{p}}}{1 + \dfrac{2L}{\pi r_{\mathrm{p}}}} \Delta c. \tag{8}$$

Introducing the resistance $R$ to colloid flow $J = \frac{\Delta c}{R}$ provides a natural interpretation of Eq. (8) in terms of the total resistance of the pore:

$$R = \frac{L}{D_0 \pi r_{\mathrm{p}}^2} + \frac{1}{2D_0 r_{\mathrm{p}}} = R^0_{\mathrm{int}} + R^0_{\mathrm{ext}}, \tag{9}$$

where the superscript '0' refers to the bare pore. The first term in Eq. (9) represents the resistance of the pore interior. The second term is the Rayleigh resistance (Eq. (7)) and accounts for the effects of convergent transport toward the pore entrance and its symmetric counterpart at the pore exit. Inside the pore, the flux lines are approximately axial. In the bare pore scenario, the inverse of the diffusion constant ($\rho_0 = D_0^{-1}$) represents the resistivity of the medium both inside and outside the pore. Naturally, the resistance for a thin membrane is determined by the exterior region ($R \approx R^0_{\mathrm{ext}}$ for $L \ll r_{\mathrm{p}}$), whereas for long pores the inner region becomes dominant ($R \approx R^0_{\mathrm{int}}$ for $L \gg r_{\mathrm{p}}$).

The finite size of colloids affects the diffusive flux in two ways. First, the excluded volume reduces the effective pore radius ($r_{\mathrm{p}} = r^0_{\mathrm{p}} - d/2$) and increases the effective pore length ($L \approx L_0 + d$) (*35*). Second, when the distance to the pore wall is comparable to the size of the colloid, some additional drag appears (*36*, *37*). We neglect the latter as the presence of the polymer brush screens hydrodynamics and thus requires a different kind of drag analysis, as discussed below.

**Internal vs. external resistances.** In the general case of position-dependent resistivity $\tilde{D}(r,z)$, we consider a set of approximate equipotential surfaces ($\psi(r,z) = \mathrm{const}$). Analogously to a set of resistors connected in parallel, the total conductivity of a layer between two adjacent equipotential surfaces is obtained by integrating local conductivity over the layer.

Inside the pore, $|z| \leq L/2$, the surfaces are taken as discs of radius $r_\mathrm{p}$ normal to the pore axis, and the conductivity is

$$\varrho_\mathrm{int}^{-1}(z) = 2\pi \int_0^{r_\mathrm{p}} \rho^{-1}(r,z) r\, dr \tag{10}$$

Outside the pore, $|z| > L/2$, we use oblate hemispheroids taken from the Rayleigh solution (*32*) (Figure 2), and

$$\begin{aligned}
\varrho_\mathrm{ext}^{-1}(z) &= 2\pi \int_0^{r_\mathrm{p}} \rho^{-1}\left(r'(r,z), z'(r,z)\right) \tilde{h}(r,z) dr \\
r'(r,z) &= r\sqrt{1 + \frac{(z-L/2)^2}{r_\mathrm{p}^2}} \\
z'(r,z) &= (|z| - L/2)\frac{\sqrt{r_\mathrm{p}^2 - r^2}}{r_\mathrm{p}} + \mathrm{sign}(z)\frac{L}{2} \\
\tilde{h}(r,z) &= h_r h_\theta h_z^{-1} = \frac{r}{r_\mathrm{p}} \frac{r_\mathrm{p}^2 + (|z| - L/2)^2}{\sqrt{r_\mathrm{p}^2 - r^2}}
\end{aligned} \tag{11}$$

where the functions $r'(r,z)$ and $z'(r,z)$ parameterize the equipotential surfaces, and $h_r$, $h_\theta$ and $h_z$ are corresponding Lamé coefficients (Supplementary Note 2). For the homogeneous brush considered here, the function $\varrho_\mathrm{ext}^{-1}(z)$ is even.

Since the consecutive layers are connected in series, their total resistance is found by integration:

$$R_\mathrm{int} = \int_{-L/2}^{+L/2} \varrho_\mathrm{int}(z) dz, \tag{12}$$

$$R_\mathrm{ext} = 2\int_{+L/2}^{+\infty} \varrho_\mathrm{ext}(z) dz \tag{13}$$

For a bare pore without a polymer brush ($\tilde{D}(r,z) = D_0$), the integrations recover Equation 9, as expected.

### Diffusivity and insertion free energy control diffusive transport

Equations 3 and 6 indicate that the flux is determined by the local diffusivity $D(r,z)$ and the insertion free energy $\Delta F(r,z)$ of the colloid, which in turn are linked to the spatial distribution of the polymer density.

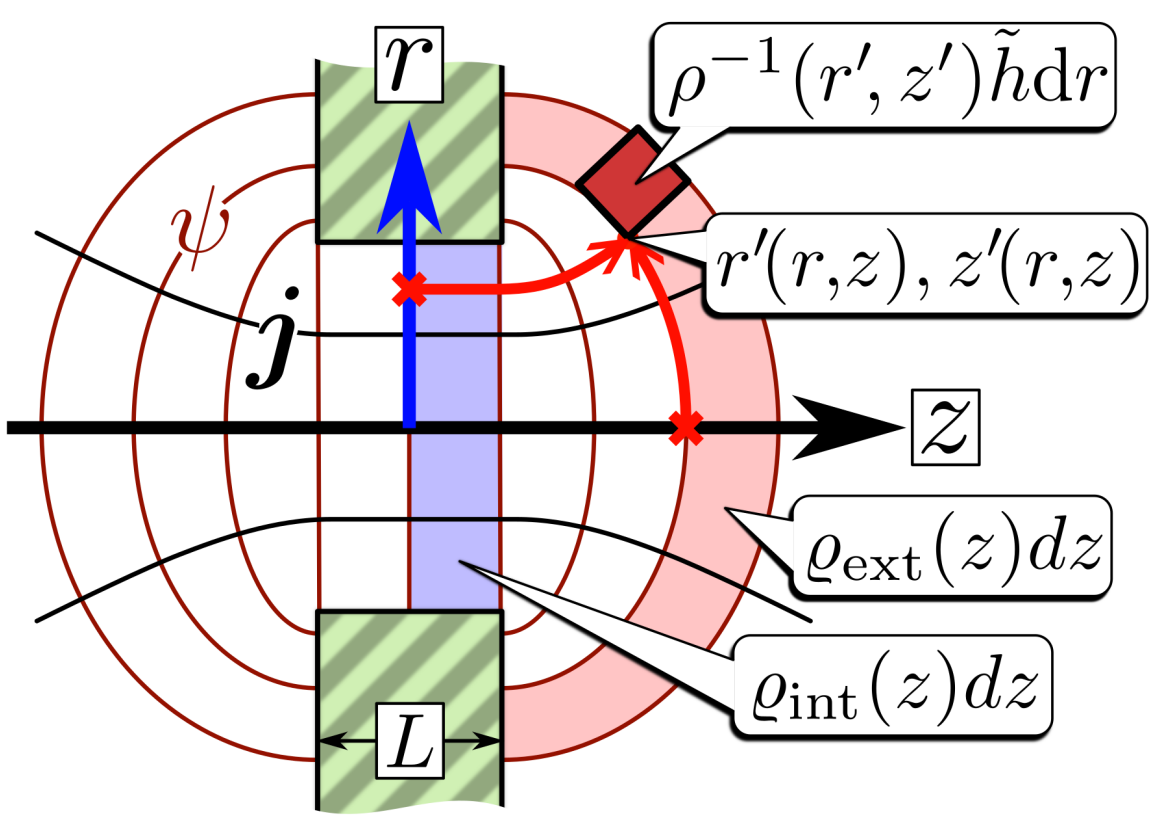

**Figure 2**: **Integration scheme for the layer resistance.** The intrinsic orthogonal curvilinear coordinates are defined by equipotential surfaces, $\psi$ = const, and the flux lines of the flux density field $\boldsymbol{j}$. A layer between two adjacent equipotential surfaces with a thickness d$z$ along the central axis has resistance $\varrho_{\mathrm{int}}\mathrm{d}z$ inside the pore ($|z| \leq L/2$, blue shading) and $\varrho_{\mathrm{ext}}\mathrm{d}z$ outside the pore (red shading). The red rectangle illustrates the local conductivity element at $(r', z')$ within an outside layer. The parametrization $(r'(r, z), z'(r, z))$ traces the integration path and maps the intrinsic coordinates back to the original cylindrical coordinates $(r, z)$, as indicated by the red arrows.

**Polymer density profiles.** Conformations adopted by overlapping polymer chains grafted to the pore walls are controlled by the solvent quality, quantified by the Flory-Huggins parameter $\chi_{\mathrm{PS}}$. Values of $\chi_{\mathrm{PS}} < 0.5$ and $\chi_{\mathrm{PS}} > 0.5$ correspond to good and poor solvent, respectively, and $\chi_{\mathrm{PS}} = 0.5$ represents the ideal (or $\theta$-)solvent.

The polymer density profile $\phi(z, r)$ in the pore was calculated by the two-gradient self-consistent field numerical method of Scheutjens and Fleer (SF-SCF; Supplementary Note 3). In Figure 3, one can appreciate the expected increase in polymer concentration inside the pore with decreasing solvent quality (increasing $\chi_{\mathrm{PS}}$). With the selected pore and polymer parameters (Figure 1), the polymer brush fills the entire pore cross-section within the full range of solvent qualities explored, and colloids need to navigate the polymer meshwork to traverse the pore. For wider pores, shorter polymers and/or lower grafting densities, an open channel free of polymer may appear in the pore center, as detailed elsewhere (*2*, *38*); this scenario would result in a different permeation behavior and is not considered here.

Figure 3 further illustrates that while the brush remains confined within the pore lumen in poor solvent ($\chi_{\rm PS} = 0.9$) it protrudes substantially into the surrounding space in ideal and good solvents ($\chi_{\rm PS} \leq 0.5$). The polymer brush will therefore affect the resistance to colloid flow outside the pore as well:

$$R = R_{\rm int} + R_{\rm ext}, \tag{14}$$

with $R_{\rm int} \rightarrow R^0_{\rm int}$ and $R_{\rm ext} \rightarrow R^0_{\rm ext}$ in the limit of the bare pore.

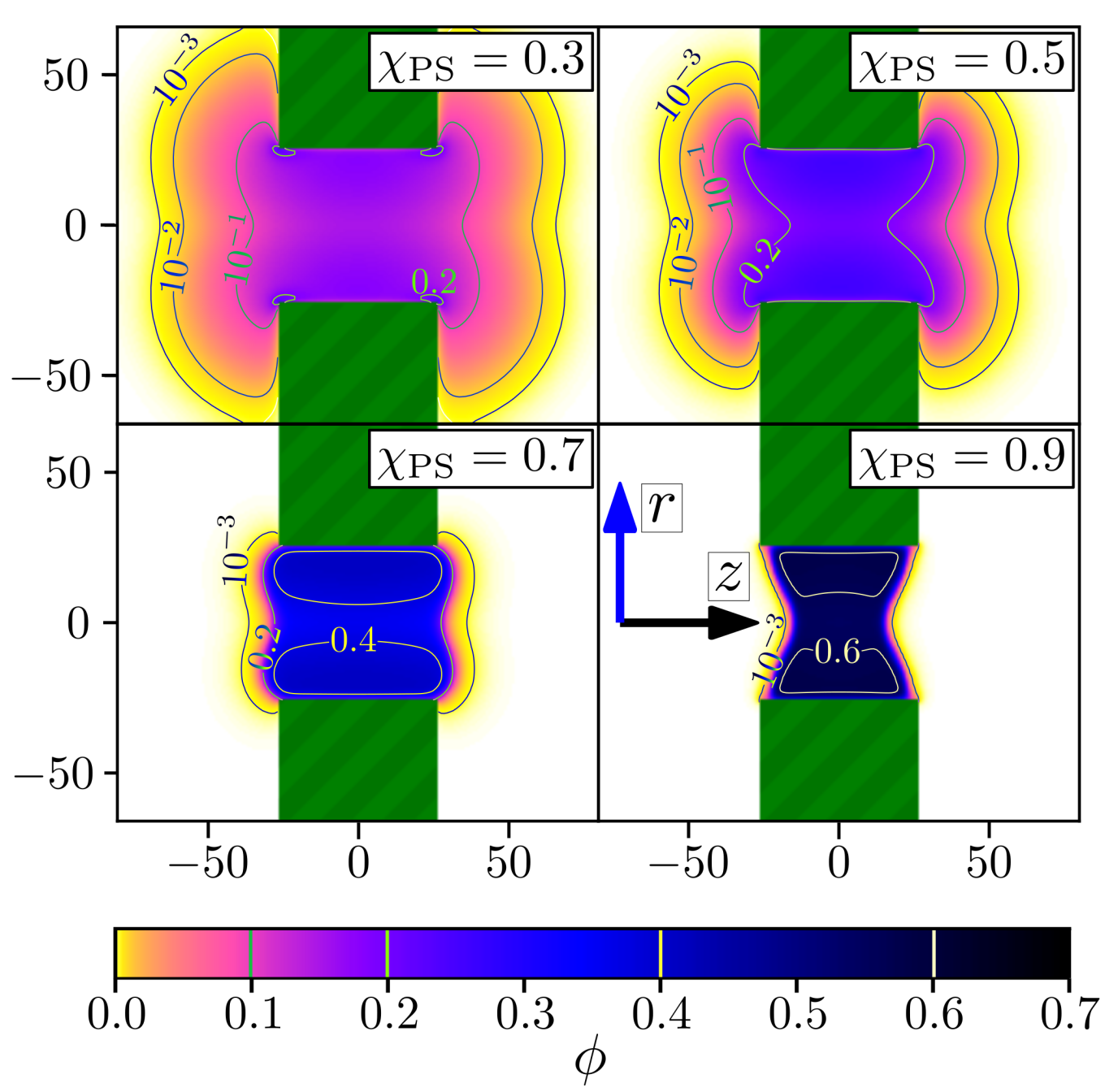

**Figure 3**: **Maps of the polymer volume fraction.** Polymer volume fraction $\phi(r, z)$ for a polymer brush in a cylindrical pore with solvent quality ranging from good (upper left panel) to poor (lower right panel), quantified by the Flory-Huggins parameter $\chi_{\mathrm{PS}}$ as indicated. Polymer volume fractions are mapped in cylindrical coordinates (as shown by $rz$-coordinate arrows), color coded as indicated and with selected iso-concentration lines. The blank space corresponds to pure solvent; the membrane is colored green. Pore and brush parameters are as given in Figure 1.

**Leading contributions to the insertion free energy.** The position-dependent insertion free energy $\Delta F(r, z)$ is the work required to move a colloid from the exterior solution into the polymer brush. For colloids that are significantly smaller than the pore, wall effects can be neglected, and the insertion free energy is determined entirely by the local polymer concentration:

$$\begin{aligned} \Delta F &= \Delta F_{\rm osm} + \Delta F_{\rm sur}, \\ \Delta F_{\rm osm}(r,z) &= \int_V \Pi(r',z')dV', \\ \Delta F_{\rm sur}(r,z) &= \oint_S \gamma(r',z')dS'. \end{aligned} \tag{15}$$

The coordinates $(r, z)$ refer to the colloid center, and the two contributions are obtained by integrating over the colloid volume $V$ and surface $S$, respectively.

The osmotic contribution, $\Delta F_{\rm osm}$, accounts for the work against excess osmotic pressure upon insertion of the colloid into the brush. The local osmotic pressure (normalized by $k_{\rm B}T$) is calculated from the local polymer concentration as

$$\begin{aligned} \Pi(r,z) = \phi(r,z)\frac{\partial f\{\phi(r,z)\}}{\partial \phi(r,z)} - f\{\phi(r,z)\} = \\ [-\ln(1-\phi(r,z)) - \phi(r,z) - \chi_{\rm PS}\phi^2(r,z)], \end{aligned} \tag{16}$$

where

$$f\{\phi(r,z)\} = (1-\phi(r,z))\ln(1-\phi(r,z)) + \chi_{\rm PS}\phi(r,z)(1-\phi(r,z))$$

is the mean-field Flory expression for the interaction free energy per unit volume of the polymer solution of concentration (volume fraction) $\phi(r, z)$.

The interfacial contribution, $\Delta F_{\rm sur}$, comprises the surface tension $\gamma(r, z)$ as the change in the free energy of a unit area of the colloid upon replacement of a contact with solvent by a polymer solution of concentration $\phi(r,z)$. $\gamma$ is approximated as

$$\gamma(r,z) = \frac{1}{6}(\chi_{\text{ads}} - \chi_{\text{ads}}^{\text{crit}})\phi^*(r,z), \tag{17}$$

$$\text{with } \chi_{\text{ads}} = \chi_{\text{PC}} - \chi_{\text{PS}}(1-\phi^*), \text{ and } \phi^*(r,z) = (b_0 + b_1\chi_{\text{PC}})\phi(r,z).$$

The coefficients $b_0 = 0.7$ and $b_1 = -0.3$ are phenomenological parameters accounting for polymer depletion or accumulation near the colloid, and can be treated as constant to a good approximation (Supplementary Note 4).

We applied an approximate analytical scheme (Supplementary Note 5) to evaluate the insertion free energy $\Delta F(r,z)$ as $\Delta F\{\phi(r,z)\}$, where $\phi(r,z)$ is the polymer density distribution in a colloid-free brush calculated with the SF-SCF approach (Supplemenatary Note 3). The colloid is thus considered as a 'probe' which does not perturb the global polymer concentration distribution $\phi(r,z)$. This scheme provided insertion free energies at any position, including off the pore axis. Comparison with direct SF-SCF calculations for colloids on the pore axis demonstrated good quantitative agreement (Supplementary Note 4), justifying the use of the more versatile analytical scheme.

**Local colloid mobility.** The crowded polymers naturally decrease the diffusion of colloids. A polymer brush is effectively described as an inhomogeneous semi-dilute polymer solution with a concentration-dependent correlation length (mesh size) $\xi(\phi)$. Colloids of size $d > \xi$ experience additional friction as they are trapped by the polymer meshwork. As a result, diffusion is slowed compared to pure solvent, leading to a position-dependent diffusion coefficient $D(r,z) < D_0$.

According to a scaling theory by Cai et al. (*39*) for diffusion of non-sticky colloids in a semi-dilute polymer solution, the colloid mobility scales as $D \sim D_0(\xi/d)^2 \ll D_0$ for $d \gg \xi$, while small colloids diffuse virtually unimpeded ($D \sim D_0$ for $d \ll \xi$). We here use a simple interpolation formula to capture the diffusion coefficient across the full range of relevant colloid sizes relative to the correlation length $d/\xi$:

$$D\{\phi(r,z)\} = \frac{D_0}{1 + [\beta d/\xi\{\phi(r,z)\}]^2} \approx \frac{D_0}{1 + [\beta d\phi(r,z)]^2}. \tag{18}$$

The correlation length $\xi$ in Eq. (18) is controlled by the local polymer concentration $\phi(r,z)$ and also depends on the solvent quality (*40*). On the right hand side we approximated $\xi \cong \phi^{-1}$, valid close to ideal solvent conditions in a mean-field regime. The coefficient $\beta$ in Eq. (18) is a numerical pre-factor. By comparing our theoretical stationary flux predictions to experimental data on the flux of non-sticky colloids of different sizes through nuclear pore complexes (see below), we estimate $\beta = 5.5$. In the following sections we consistently use this value for numerical estimates of the pore permeability and permselectivity.

Recent theoretical, experimental and simulation studies extended the scaling approach of Cai et al. (*39*) to include the effects of polymer-colloid attraction (*41*, *42*). In the limit where proper diffusion of polymer chains is negligible (relevant to our case of pore-anchored chains), the characteristic desorption time of a single polymer-colloid contact $\tau_{\mathrm{des}}(\epsilon)$ appears as an extra timescale. Here, $\epsilon$ is the absolute value of the activation energy to break a contact. The polymer-colloid attraction may be considered weak if the desorption time is much smaller than the self-diffusion time. Typically, this happens when $\epsilon \lesssim k_{\mathrm{B}}T$, and in this case attractive forces only modify local friction by affecting the local packing of polymers around the colloid. The magnitude of this effect is estimated to be small, reducing the diffusivity by up to a few 10% (*43*). On the other hand, the desorption time grows exponentially with $\epsilon$ and affects the diffusion substantively when $\epsilon$ amounts to several $k_{\mathrm{B}}T$ units. Numerical simulations demonstrated that the colloid diffusivity is reduced by (*41*)

$$D(\epsilon) \approx D(\epsilon = 0) \exp(-\epsilon/2). \tag{19}$$

In our model, the energy of a polymer-colloid contact is $\chi_{\mathrm{PC}}/6$, and therefore $\epsilon < 1/3$ within the explored range $0 \geq \chi_{\mathrm{PC}} \geq -2$. Throughout most of the paper, we assume that the surface of the colloid is homogeneous, and neglect the small reduction in the local colloid mobility due to weak polymer attraction. However, we will return to this question in the context of proteins traversing the nuclear pore, where the colloid surface and the polymers may be heterogeneous with affinity localized in a few sticky patches. We can explore the limiting case where all the (negative) surface free energy $\pi d^2 \gamma(r, z)$ is assigned to a single sticky patch. The diffusion coefficient is then modified as

$$D_{\text{sticky patch}} = \begin{cases} D \exp(\frac{\pi d^2 \gamma(r,z)}{2}) & \text{if } \gamma(r, z) < 0 \\ D & \text{otherwise} \end{cases} \tag{20}$$

to interpolate between non-sticky and sticky colloids. It is safe to assume that realistic cases of attractive polymer-colloid interaction lie between the two extremes of a perfectly homogeneous surface ($D_{\rm homo} \approx D$) and a single sticky patch (Eq. (20)).

Several other theoretical and empirical models have been proposed to describe the diffusion of colloids in polymer meshworks (*44–47*). Although the predictions of different models differ quantitatively, they all share the same qualitative trend.

**Resistivity maps derived from polymer density.** Following Eq. (6), the local resistivity anywhere in the polymer phase can be expressed as the sum of diffusivity and insertion free energy contributions, $\ln[\rho(\phi)D_0] = -\ln[D(\phi)/D_0] + \Delta F(\phi)$. Figure 4a illustrates the distinct dependencies of these two contributions on the polymer volume fraction: while $-\ln[D(\phi)/D_0]$ increases monotonically (black line), $\Delta F(\phi)$ can be non-monotonic (colored lines). For sufficiently attractive colloids ($\chi_{\rm PC} < \chi_{\rm ads}^{\rm crit} + \chi_{\rm PS}$), $\Delta F(\phi)$ transits from an attractive ($\Delta F < 0$) to a repulsive ($\Delta F > 0$) regime at a finite $\phi$ where the interfacial gain $\Delta F_{\rm sur}$ balances the osmotic penalty $\Delta F_{\rm osm}$. Inert or weakly attractive colloids ($\chi_{\rm PC} \geq -0.5$) always have $\Delta F > 0$ and are thus repelled.

Figure 4b-c illustrates the spatial distribution of the diffusivity and free energy contributions, respectively, for a colloid of size $d = 8$ in an in an ideal solvent ($\chi_{\rm PS} = 0.5$) and a set of polymer-colloid interaction strengths ($\chi_{\rm PC}$). Because the polymer concentration is highly inhomogeneous, $\Delta F(r, z)$ varies strongly: for example, at $\chi_{\rm PC} = -0.75$ the pore fringes are attractive while the dense interior is repulsive. Thus, diffusivity and free energy, and consequently the local resistivity, are spatially resolved functions of the polymer volume fraction.

### An attractive polymer filling enhances colloid fluxes through the pore

Figure 5 visualizes the relative contributions of the pore interior ($R_{\rm int}$) and exterior ($R_{\rm ext}$) to the total resistance as a function of the polymer-colloid attraction $\chi_{\rm PC}$, for a selected colloid size ($d = 12$) in a good ($\chi_{\rm PS} = 0.3$) and a poor ($\chi_{\rm PS} = 0.7$) solvent. A comparison of the total resistance obtained by direct numerical solution of the Smoluchowski equation (diamond symbols; Supplementary Note 6) here validates the accuracy of the analytical scheme (solid lines).

A striking feature is that attracted colloids can achieve diffusive fluxes that exceed the limit of the bare pore, as indicated by the segments of the $R(\chi_{\rm PC})$ curves that fall below the solid horizontal red line marking the bare-pore resistance $R_0$. This result may at first appear surprising, given that the polymer medium is expected to slow down the diffusion of colloids. However, this slowing down is counteracted by the attractive potential of the polymer meshwork, which reduces local resistivity according to the exponential factor in Eq. (6).

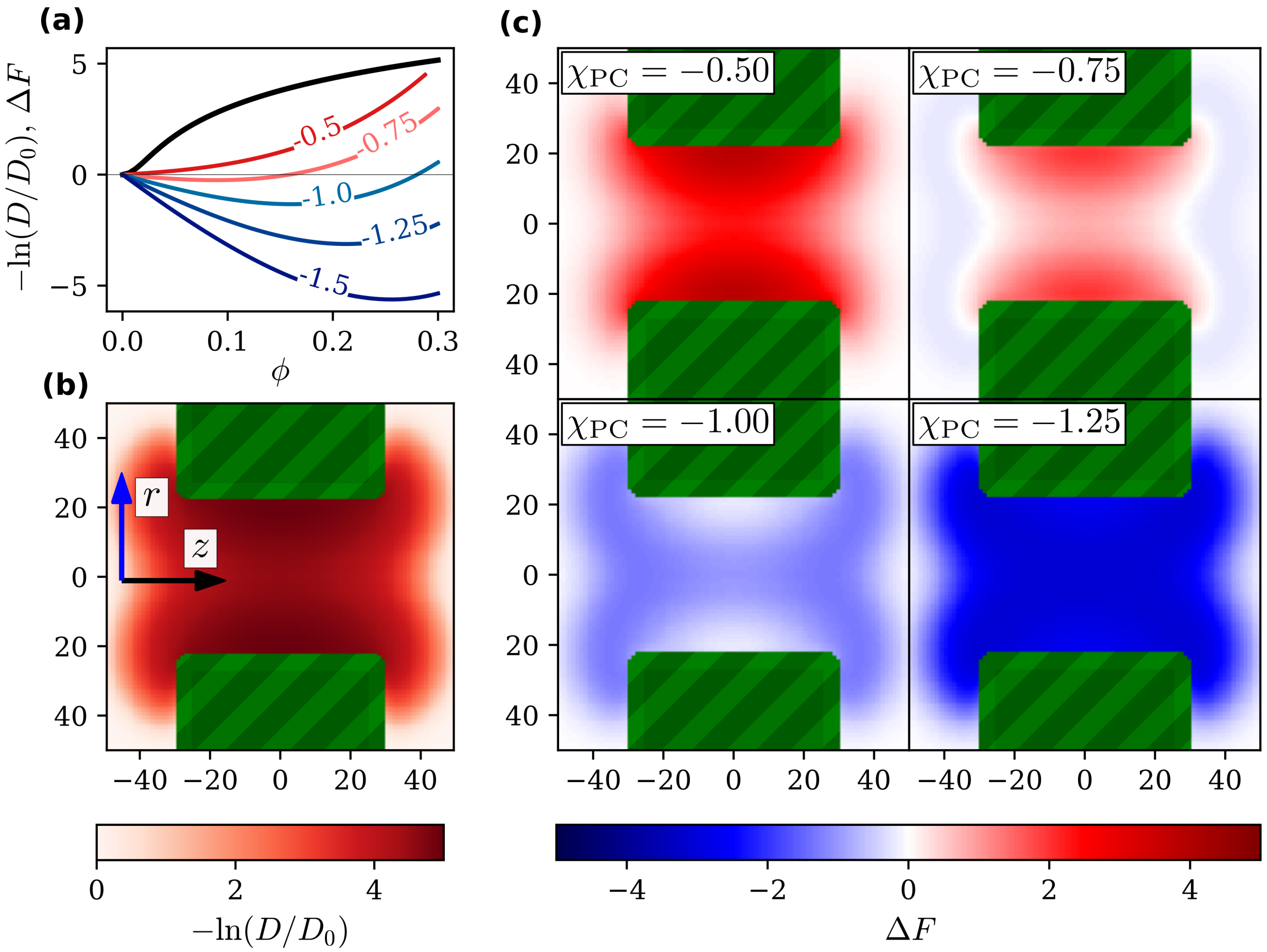

**Figure 4**: **Effect of the polymer meshwork on local diffusivity and insertion free energy. (a)** Comparison of the reduced colloid diffusivity caused by the polymer meshwork ($-\ln(D/D_0)$; black line) with the insertion free energy (colored lines, labelled by $\chi_{PC}$) as functions of the local polymer volume fraction $\phi$. Both quantities share the same vertical scale so that the sum of the black and one of the colored curves, $\ln(\rho D_0) = -\ln(D/D_0) + \Delta F$, quantifies the total local resistivity; the horizontal zero line corresponds to unimpeded transport as in pure solvent; positive values of the colored curves indicate resistance enhanced by the free energy contribution, whereas negative values indicate lowered resistance. **(b)** Spatial map of the position-dependent diffusivity, $-\ln\left[D(\phi(r,z))/D_0\right]$, in cylindrical coordinates. **(c)** Maps of the insertion free energy $\Delta F(r,z)$ for polymer-colloid interaction strengths ranging from $\chi_{PC} = -0.50$ (weakest attraction) to $\chi_{PC} = -1.25$ (strongest attraction), as indicated. Pore and brush parameters are as in Figure 1; $\chi_{PS} = 0.5$ and $d = 8$.

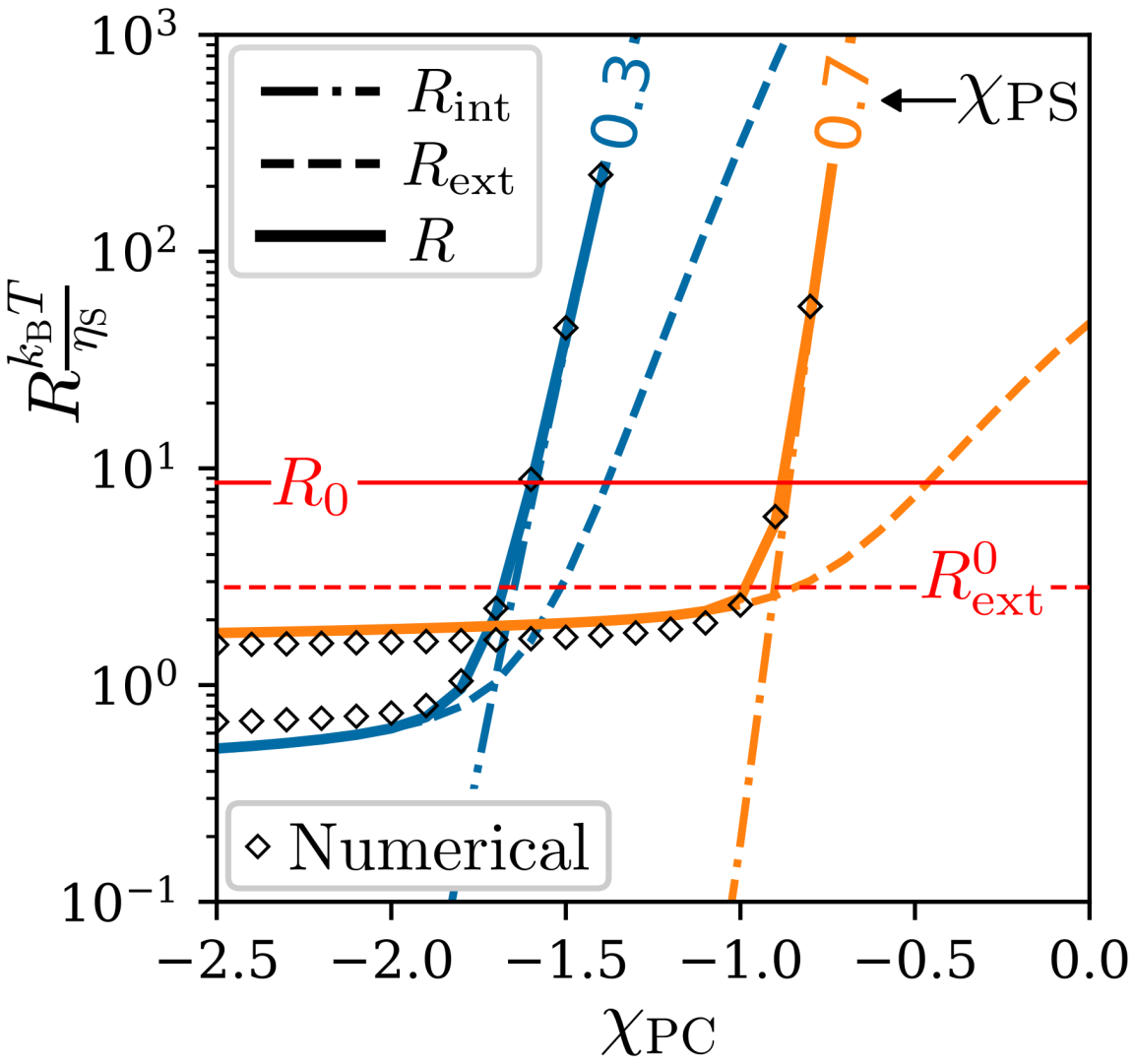

**Figure 5**: **Resistance contributions vs. polymer-colloid interaction strength.** Interior ($R_{\mathrm{int}}$), exterior ($R_{\mathrm{ext}}$) and total ($R = R_{\mathrm{int}} + R_{\mathrm{ext}}$) resistance vs. polymer-colloid interaction strength $\chi_{\mathrm{PC}}$ for a good solvent ($\chi_{\mathrm{PS}} = 0.3$, in blue) and a poor solvent ($\chi_{\mathrm{PS}} = 0.7$, in orange), obtained throught the analytical scheme. The exterior $R^0_{\mathrm{ext}}$ and total $R_0$ resistances for the bare pore are also shown (red lines of matching type). The resistances are presented in dimensionless units $R\frac{k_BT}{\eta_S}$; pore and brush parameters are as given in Figure 1; $d = 12$. The results of direct numerical solution of the Smoluchowski equation are shown with diamond symbols.

The reduced local resistivity has pronounced consequences for diffusive transport in both the interior and the exterior of the pore. Figure 5 illustrates that the interior resistance $R_{\mathrm{int}}$ can be driven practically to zero by increasing the polymer-colloid attraction (decreasing $\chi_{\mathrm{PC}}$) below a certain threshold. Compared to a bare pore (Eq. (9)) such a short-circuiting effect entails a reduction in the resistance by a factor of up to $R^0_{\mathrm{int}}/R^0_{\mathrm{ext}} + 1 \approx 2/\pi \times L/r_{\mathrm{p}} + 1$. For the pore and colloid considered here, this represents an approximately 3-fold reduction, to a level marked by the dashed horizontal red line in Figure 5 which equals the contribution of convergent flux to the bare-pore resistance ($R^0_{\mathrm{ext}}$). The reduction would be even stronger for longer pores ($L \gg r_p$), and for larger colloids that increase the effective pore length and decrease the effective pore diameter.

The exterior resistance $R_{ext}$, on the other hand, always retains a finite contribution from the diffusive fluxes in the semi-infinite reservoir, setting an absolute lower bound to the total resistance. The reduction of $R_{ext}$ below $R^0_{ext}$ evidenced in Figure 5 is due to attractive brush fringes that protrude and facilitate diffusive transport outside the pore. Approximating the brush fringes on either end of the pore as hemispherical caps with radius $r_{cap}$, the plateau conditions are equivalent to the resistance of an ideally absorbing sphere (*48*),

$$R^{min}_{ext} = 1/(D_0 \pi r_{cap}). \tag{21}$$

Attractive brush fringes thus entail a reduction in resistance by a factor of up to $R^0_{ext}/R^{min}_{ext} = \pi/2 \times r_{cap}/r_p$. In good solvent ($\chi_{PC} = 0.3$), for example, the cap radius (along the pore axis) is comparable to the pore diameter (Figure 3), leading to a 3.3-fold reduction of the external resistance, and a cumulative 10-fold reduction of the total resistance, compared to the bare pore (Figure 5). The cap size shrinks as the solvent quality decreases (Figure 3), with a correspondingly reduced benefit on pore conductivity, as illustrated for $\chi_{PC} = 0.7$. For even poorer solvents, the cap and its benefit disappear entirely ($R_{ext} = R^0_{ext}$; not shown).

**Polymer-filled mesopores effectively gate colloids by their attraction to the polymer**

Figure 5 also illustrates how the total resistance of the pore varies with the colloid's affinity to the polymer brush. As expected, increasing the polymer-colloid attraction strength (i.e., more negative $\chi_{PC}$) results in a monotonic decrease in the pore's total resistance, since the interfacial term in the insertion free energy becomes more negative, thereby increasing the local conductivity $\rho^{-1}$.

Most notable is a sharp transition from a regime of facilitated permeation ($R < R_0$) to a regime of impeded permeation ($R > R_0$). The regime of impeded permeation is dominated by the internal resistance. It exhibits high selectivity with respect to the polymer-colloid interaction strength, and a mostly very high total resistance and thus low colloid flux, both appreciable in Figure 5 as a sharp increase in $R$ over a relatively modest $\chi_{PC}$ range. In contrast, the region of facilitated permeation is dominated by the external resistance. It exhibits high colloid fluxes but rather low (if any) $\chi_{PC}$ selectivity, as demonstrated by the previously analyzed plateau. Thus, the transition between the two regimes of transport defines the condition for sharp colloid gating, with remarkably efficient transport in the regime limited by external resistance and effective blockage in the regime limited by internal resistance.

In this context, the solvent quality can be seen as a regulator of the polymer-colloid interaction level for gating. Lowering the solvent quality (increasing $\chi_{\mathrm{PS}}$) reduces $\chi_{\mathrm{ads}}$ and shifts the entire $R(\chi_{\mathrm{PC}})$ curve toward larger $\chi_{\mathrm{PC}}$ values. Thus, a poorer solvent extends the range of facilitated permeation towards more weakly interacting colloids.

The here-presented trends are qualitatively correct also for colloids with sizes smaller or larger than the $d = 12$ considered here. Naturally, the gating effect will be rather moderate for small colloids, yet even sharper for larger colloids.

**High colloid flux implies colloid enrichment in the pore**

Colloid concentration profiles under stationary flux conditions can be found by numerically solving Eq. (1) with $\frac{\partial c(r,z)}{\partial t} = 0$ (Supplementary Note 6). Figure 6 maps the steady-state colloid concentration across a polymer-filled pore for a colloid of size $d = 8$ and polymer-colloid interaction strength $\chi\mathrm{PC} = -1.25$ in an ideal solvent ($\chi\mathrm{PS} = 0.5$). This condition corresponds to transport rates somewhat inferior to the bare pore ($R \approx 5R_0$), compared to same-size but inert colloids where the resistance increases by orders of magnitude ($R(\chi_{\mathrm{PC}} = 0) \approx 10^4 R_0$).

The map illustrates several salient features of the diffusion process. Inside the pore, the flux lines remain nearly parallel to the pore axis. Outside the pore and polymer fringes, the concentration rapidly approaches the bulk value of each semi-infinite reservoir. Also, the corresponding map of the potential $\psi$ (Figure 6, inset) reveals the same shape of equipotential surfaces as the bare pore (e.g., oblate hemispheroids), further validating the analytical scheme used earlier to evaluate the total resistance.

The most notable observation is that the colloid concentrations substantively exceed $c_0$ near the pore entrance (by a factor of $\sim 20$) and inside the pore (by a factor of $\sim 10$). This effect is caused by the negative insertion free energy in the space occupied by the polymer brush. At equilibrium (i.e., with vanishing fluxes), the partitioning would amount to $c_{\mathrm{eq}}/c_0 = \exp(-\Delta F)$. In the steady state (i.e., with non-vanishing fluxes), the colloid concentration is reduced but approaches the equilibrium concentration as the insertion free energy becomes very negative ($c/c_0 \to c_{\mathrm{eq}}/c_0$).

The presented quantitative results are only valid for sufficiently low bulk concentrations $c_0$, as our model disregards any colloid crowding effects. Such interactions are expected to be predominantly of excluded-volume (repulsive) nature, which would systematically lower the steady-state colloid concentration inside the brush.

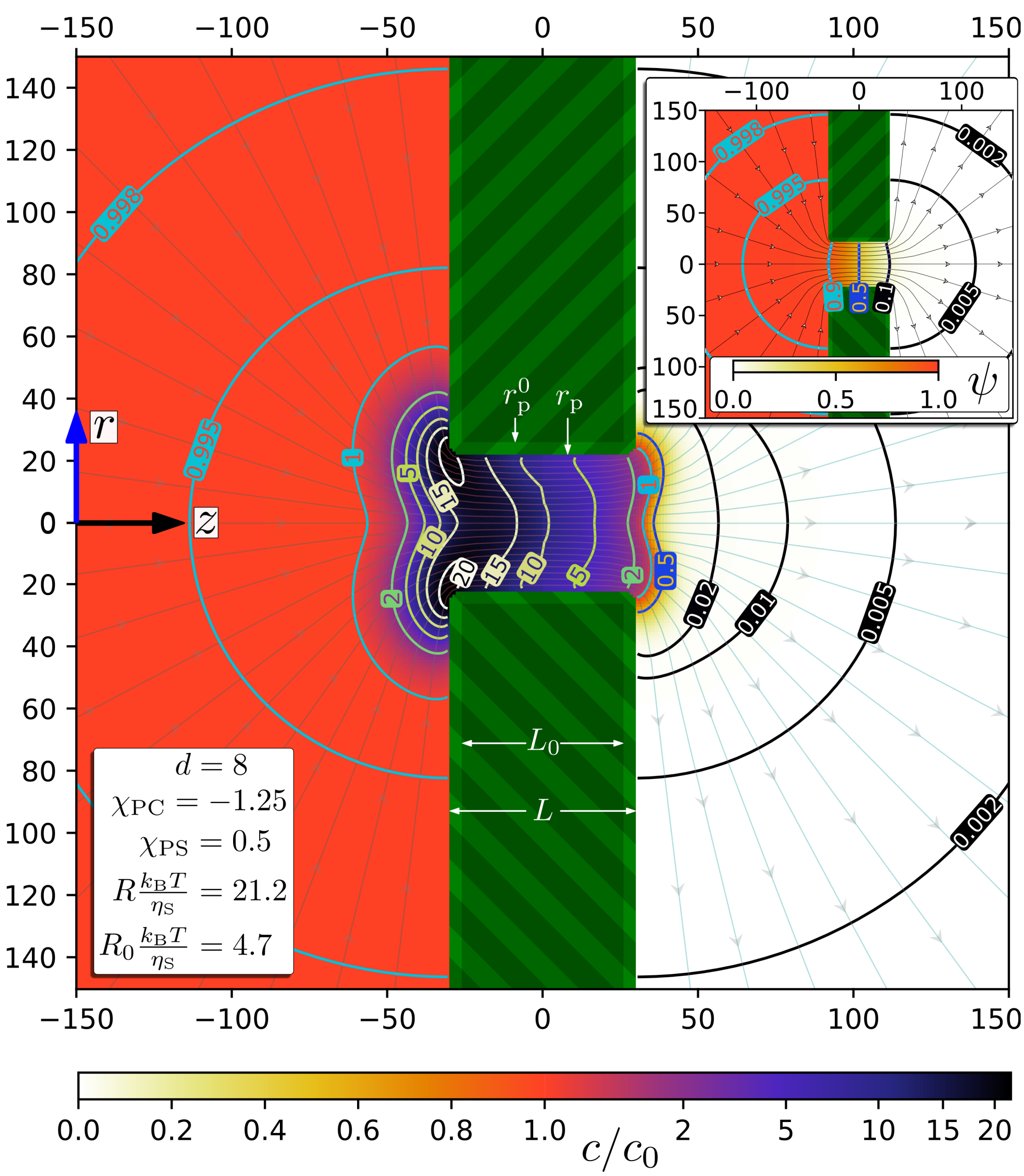

**Figure 6**: **Steady-state colloid concentration.** The colormap of the steady-state colloid concentration is normalized by the bulk concentration in the source compartment $c_0$. Isoconcentration contours are shown as labeled. The flux is represented by streamlines marked with small arrows, indicating the average colloid trajectory. Pore and brush parameters are the same as in Figure 1; $d = 8$, $\chi_{\mathrm{PC}} = -1.25$ and $\chi_{\mathrm{PS}} = 0.5$. The physical pore radius ($r_{\mathrm{p}}^0$) and length ($L_0$), along with their effective counterparts due to colloid excluded volume ($r_{\mathrm{p}}$ and $L$, respectively), are shown with white arrows. The inset (top right) shows a part of the corresponding $\psi$ potential map.

### Polymer-filled mesopores effectively gate colloids by their size

Figure 7 compares how the total resistance, $R = R_{\mathrm{int}} + R_{\mathrm{ext}}$, varies with colloid size $d$ for a bare pore (thick black lines) and for polymer-filled pores with selected solvent (Figure 7a) and polymer-colloid interaction (Figure 7b) strengths (thin colored lines with symbols). For small colloids, the bare-pore resistance follows well the $R_0 \sim D_0^{-1} \sim d$ dependence expected according to Eq. (9). A stronger dependence of $R_0$ on $d$ observed for larger colloids is due to decreasing effective pore length $L$ and increasing effective pore radius $r_{\mathrm{p}}$.

Naturally, the polymer filling affects the transport of the smallest colloids only marginally, as their volume and net interaction strengths (within the considered $\chi_{\mathrm{PC}}$ range) are too small to have any noticeable effect. A rich picture emerges for larger colloids, however, with non-monotonic dependencies of the pore resistance on colloid size and strong effects of $\chi_{\mathrm{PC}}$ and $\chi_{\mathrm{PS}}$.

**Impact of colloid diffusivity within the polymer brush on size-selective transport.** The curve with $\chi_{\mathrm{PS}} = 0.5$ and $\chi_{\mathrm{PC}} = -1.0$ in Figure 7a corresponds to the condition of $\Delta F$ fairly vanishing across a wide colloid-size range (Supplementary Note 4 - Figure S4) due to compensation of osmotic and surface contributions to the insertion free energy. Here, the reduced colloid diffusivity within the polymer meshwork dominates the pore resistance (Eq. (6)). This effect alone leads to a monotonic and pronounced increase of $R$ with $d$ with smooth crossover between asymptotic dependencies $R \sim d$ at $d \ll \xi$ and $R \sim d^3$ at $d \gg \xi$. Notably, for small and intermediate size colloids the pore resistance slightly grows with inferior solvent quality (increase in $\chi_{PS}$) due to a decrease in the mesh size $\xi$ with concomitant decrease in the local diffusivity, as seen in Figure 7b.

**Impact of insertion free energy on size-selective transport.** Since the pore resistance scales exponentially with the insertion free energy ($R \sim D^{-1} \exp(\Delta F)$; Eq. (6)), and $\Delta F = \Delta F_{\mathrm{osm}} + \Delta F_{\mathrm{sur}}$, the dependence of the resistance on colloid size is generally controlled by the interplay between the osmotic $\Delta F_{\mathrm{osm}} \sim \Pi d^3$ and the interfacial $\Delta F_{\mathrm{sur}} \sim \gamma d^2$ contributions. While the osmotic repulsion arising due to the polymer filling always enhances the resistance, the surface contribution may either increase (at $\gamma > 0$) or decrease (at $\gamma < 0$) it.

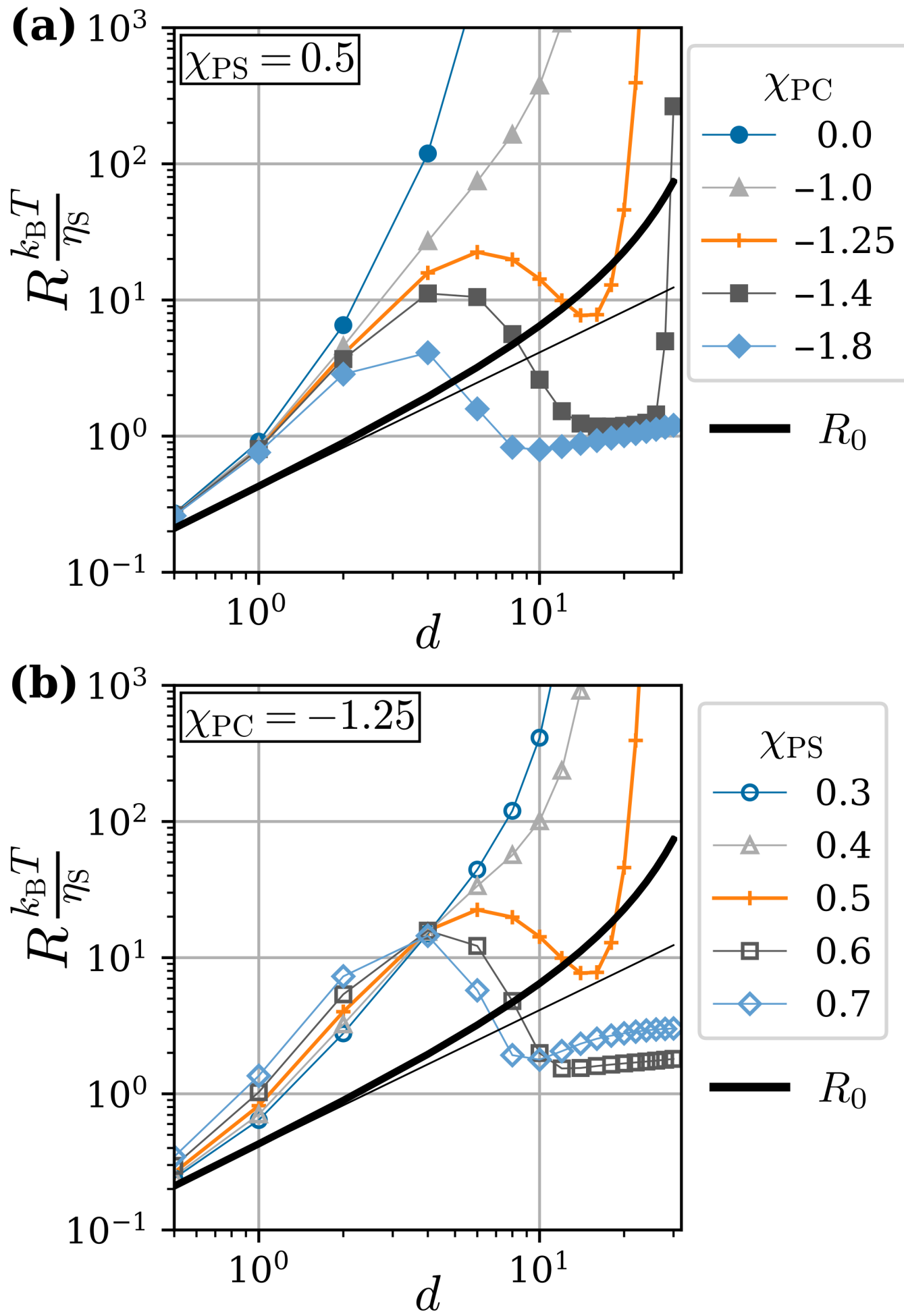

**Figure 7**: **Pore resistance as a function of the colloid size.** Normalized dimensionless pore resistance $R\frac{k_BT}{\eta_S}$ as a function of the colloid size $d$ for **(a)** selected values of the polymer-colloid interaction strength $\chi_{PC}$ (as indicated in the legend) at a fixed solvent strength $\chi_{PS} = 0.5$, and **(b)** selected values of $\chi_{PS}$ (as indicated in the legend) at a fixed $\chi_{PC} = -1.25$. Pore and brush parameters are as given in Figure 1. The bare-pore resistance $R_0$ is shown by the thick black line. Its deviation from simple Stokesian scaling (Eq. (9); thin black line) is due to the excluded volume of the colloid.

For inert or weakly attractive colloids, $\gamma \geq 0$, the resistance grows monotonically with the colloid size due to the combined effect of a decreasing diffusivity $D(d)$ and an increasing insertion free energy $\Delta F(d)$. As both these effects are pronounced, their combination leads to a very strong size selectivity, such that the transport of even rather small colloids is effectively impeded, as can be appreciated for $\chi_{PC} > -1.0$ in Figure 7a. For sufficiently large colloids, the osmotic contribution dominates in the insertion free energy such that $R \sim D^{-1} \exp(\Delta F) \sim d^3 \exp(\Pi d^3)$.

In contrast, for attractive colloids with $\gamma < 0$ the dependence of the pore resistance $R(d)$ on colloid size can be non-monotonic, with a local maximum (at $d = d_{\text{max}}$) followed by a local minimum (at $d = d_{\text{min}}$). This is best illustrated in Figure 7 by the orange curves corresponding to $\chi_{PC} = -1.25$ and $\chi_{PS} = 0.5$. The local maximum here arises from the net colloid attraction (which decreases resistance) overcoming the decrease in diffusivity (which increases resistance) with increasing colloid size. The local minimum in turn arises from the positive osmotic contribution to the free energy ($\Delta F_{\text{osm}} \sim \Pi d^3$) overcoming the negative interfacial contribution ($\Delta F_{\text{sur}} \sim \gamma d^2$).

Most notably, the local minimum for attractive colloids is swiftly followed by a sharp increase in resistance (for $d > d_{\text{min}}$) due to the dominant osmotic contribution recovering the strong $R \sim d^3 \exp(\Pi d^3)$ dependence. The transition between the regime of good to moderate transport (for $d \lesssim d_{\text{min}}$) and the regime of impeded transport (for $d > d_{\text{min}}$) thus defines the condition for sharp gating of attracted colloids by their size.

When the insertion free energy becomes strongly negative, a new regime appears that is characterized by facilitated transport ($R < R_0$) over a rather wide range of colloid sizes, as illustrated in Figure 7a for $\chi_{PC} \leq -1.4$, and in Figure 7b for $\chi_{PS} \geq 0.6$. Here, the pore interior is effectively short-circuited, $R_{\text{int}} \to 0$, and the total resistance is set by the finite exterior contribution, $R \approx R_{\text{ext}}$ (see Eq. (14)). Due to attractive brush fringes at the pore entrance and exit, $R \approx R_{\text{ext}}^{\text{min}}$, leading to a weak size dependence, $R \sim d$ (see Eq. (21)).

The sharp gating of colloids by their size is preserved, and even enhanced, for strongly attractive colloids. This is best seen in Figure 7a for $\chi_{PC} = -1.4$, where the osmotic penalty to the insertion free energy takes over, and entails a sharp increase in resistance, above a certain colloid size ($d \approx 24$).

**Experiments of colloid transport through NPCs validate the theoretical predictions**

To test our theory, we analyzed literature pertinent to colloid transport across NPCs. The average distance between NPCs in the nuclear envelope exceeds the pore diameter by almost an order of magnitude (*49–51*). Transport across neighbouring NPCs thus is not mutually interfering (*52*) (see iso-concentration lines in Figure 6). Experimentally measured transport rates $k$, normalized against the number of pores (Supplementary Note 7), therefore can be directly compared with our theoretical predictions.

It is well-known that colloids with affinity for the disordered nucleoporin FG domains that fill the NPC (such as importins and exportins) are enriched in or near NPCs (*18*, *53*, *54*), and in microscopic droplets, macroscopic hydrogels and thin films assembled from pure FG domains (*21*, *23–25*). Qualitatively, these observations fully align with our predictions that the accumulation of colloids in the pore is required for facilitated transport (Figure 6). We hence tested our predictions quantitatively.

**Transport of non-sticky colloids.** Literature on the rates of diffusive transport across NPCs for non-sticky proteins (*25–29*) collectively covers two orders of magnitude in molecular mass and five orders of magnitude in transport rate. Figure 8a compares these experimental data with the theoretical predictions of our model. As the example pore geometry and polymer density in Figure 1 were modelled to approximate NPCs (Supplementary Note 1), we can directly compare our theoretical predictions with experimental data. The effective statistical segment length of disordered polypeptide chains was taken to be $a = 0.76$ nm (*19*). The effective solvent strength in the NPC was estimated to be close to ideal ($\chi_{\mathrm{PS}} = 0.6$), consistent with varying yet generally moderate levels of 'cohesiveness' observed for FG domains (*19*, *23*, *24*, *55*, *56*). We approximated the proteins as perfectly inert colloids ($\chi_{\mathrm{PC}} = 0$). To match the theoretical colloid volumes to protein molecular masses, we considered the effective density of the protein colloids to be bounded by the densities of aqueous solvent ($\rho_{\mathrm{probe}} \geq 1$ g/cm$^3$) and pure polypeptide ($\rho_{\mathrm{probe}} \lesssim 1.4$ g/cm$^3$). This reflects that an unknown (and possibly variable) amount of solvent contributes to the effective volume of the proteins during their transport across the NPC. The only adjustable fitting parameter in our model was the prefactor $\beta$ in the scaling-based expression for the diffusion coefficient, Eq. (18).

Figure 8a demonstrates that the theory reproduces the experimentally observed increase in pore resistance with colloid size very well. The best fit was obtained with $\beta = 5.5$ and this value was hence fixed throughout the paper. The quality of the fit is quite remarkable given the large range of masses and transport rates covered, and the relative simplicity of our theory. Some scatter in the experimental data is though notable. This may be due to some proteins not being strictly non-sticky but interacting weakly with FG domains. Indeed, Frey et al. (*25*) reported a three-fold enhanced transport rate of green fluorescent protein over mCherry despite these proteins being of similar molecular mass and considered inert. Moreover, whilst some studies had washed out cytosolic proteins in their assay (with HeLa cells), thus leaving behind intact nuclear pores filled with a plain FG domain brush but lacking most transport factors (*25–27*), others used intact yeast cells with all transport factors present (*28*, *29*). The satisfactory fit across all datasets suggests that the crowding of the NPC with transport factors has a comparatively weak effect on the transport of non-sticky proteins across the NPC, consistent with a moderate effect of transport factor depletion observed in vivo (*57*).

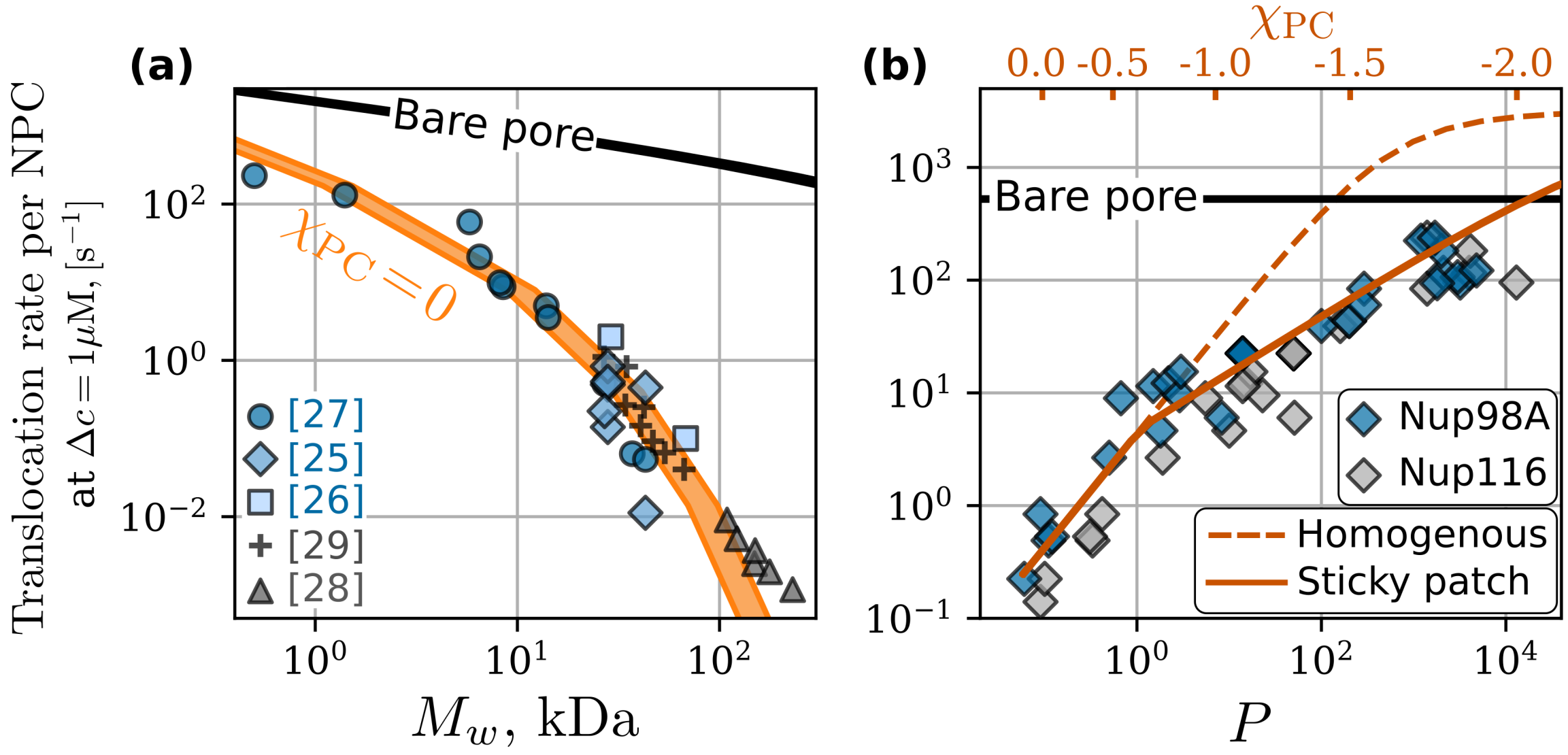

**Figure 8**: **Comparison of theoretical predictions with experimental findings for colloid transport rates across NPCs. (a)** Gating of non-sticky colloids (peptides and globular proteins) by size. NPC passage rate $k$ per pore (at $c_0$ = 1 $\mu$M) vs. molecular mass $M_w$ (symbols) extracted from the literature, as indicated (Table S1). Theoretical predictions (shaded orange area) are for the pore and brush parameters as given in Figure 1, with $\chi_{PC}$ = 0, $a$ = 0.76 nm, $\chi_{PS}$ = 0.6, and colloid volumes converted to masses using 1 g/cm$^3$ $\leq \rho_{probe} \leq$ 1.4 g/cm$^3$. The best fit, shown here, was obtained with $\beta$ = 5.5 in Eq. (18). The predicted transport rates across a bare pore are shown (black line) for comparison. **(b)** Gating of colloids by their affinity to the polymer. NPC passage rate $k$ vs. partition coefficient $P$ in phase-separated droplets of pure FG domains (Nup98A - blue lozenges, Nup116 - gray lozenges) measured by Frey et al. (*25*) for a range of green fluorescent protein variants and mCherry (Table S2). Theoretical predictions (orange lines) represent the limits of a homogeneously attractive colloid surface (dashed line) and a surface with a single sticky patch (solid line). The predicted transport rate across a bare pore is also shown (black line). The pore and brush parameters are as given in (a); $\chi_{PS}$ = 0.6 for both pore and pure FG domain droplets; $\chi_{PC}$ values (indicated at the top of the graph) were matched to the partition coefficient $P$, and $d$ = 6 (Supplementary Note 7).

**Transport of sticky colloids.** Frey et al. (*25*) additionally quantified NPC transport rates for a wide range of green fluorescent proteins (GFPs) with surface amino acids mutated to modulate transport from 'superinert' to 'transport factor like'. In parallel, the ability of these variants to enrich or deplete in phase-separated droplets of two pure FG domains (Nup98A from *T. thermophila*, and Nup116 from *S. cerevisiae*) was quantified. The transport rate was observed to correlate strongly with the level of GFP enrichment in FG domain phases (Figure 8b). This set of experiments enabled the effect of polymer-colloid interaction to be tested selectively as the colloid size and shape were effectively constant.

To reproduce theoretically the correlation between the experimentally measured NPC transport rates $k$ and the partition coefficient $P$ in pure FG domain phases, we assumed an effective solvent strength $\chi_{\mathrm{PS}} = 0.6$ for the two pure FG domain phases and the NPC. This simplified assumption has only a moderate impact on the predictions (Supplementary Note 7). Furthermore, the free energy of insertion $\Delta F = -\ln(P)$ is reduced to the surface contribution ($\Delta F = \Delta F_{\mathrm{sur}}\left(\chi_{\mathrm{PS}}, \chi_{\mathrm{PC}}\right)$) since the osmotic pressure vanishes on spontaneous phase separation at low polymer concentration (Supplementary Note 7), and provides the link between $\chi_{\mathrm{PC}}$ and the partition coefficient.

Our idealised assumption of colloids being homogeneously interactive (Figure 8b, dashed orange line) reproduced the experimental data for non-sticky and weakly attractive colloids well without any adjustable parameter. For more strongly attractive colloids however, this approach overestimated the experimentally observed transport rates. Assuming the opposite extreme of all surface free energy being concentrated into a single sticky patch (Figure 8b, solid orange line) reproduced the experimental data quite well, suggesting that the presence of localized sticky patches on the colloid surface and the FG domains slows down diffusion and transport.

Taken together, the quantitative agreement between our theory and a range of experimental data for NPC passage of proteins with a very limited number of adjustable parameters provides strong validation for our theory.

## Discussion

We have shown how mesopores filled with polymer brushes can gate transport with exquisite selectivity with respect to polymer-colloid affinity and colloid size, even for colloids that are substantially smaller than the pore diameter. A striking finding is that an attractive polymer brush can provide colloid transport rates comparable to, or even exceeding, the bare pore. Our findings shed light on the basic mechanisms of selective nucleo-cytoplasmic transport and suggest a molecular design strategy for controlling selective permeability through artificial mesoporous membranes.

### Implications for nuclear pore permselectivty

Figure 8b indicates that the theoretical limit for the rate of transport of sticky colloids is higher than what may be realised with proteins in nuclear pores. This finding is intriguing, as it suggests the rate of transport is not the primary performance factor for NPCs. Arguably, selectivity of transport may be the more important criterion, and the limited biochemical space available for nature to evolve towards high selectivity (whilst maintaining basic properties such as colloidal stability in the cellular milieu) may have come with a tradeoff in terms of rate.

Our model predicts that the highest transport rates are achieved with homogeneously attractive polymers and colloids. In contrast, each FG domain polymer type exhibits substantive heterogeneity along the chain contour with preferred interaction sites for transport factors. Similarly, importins, exportins and their cargo display substantive surface heterogeneity and complex, non-spherical shapes. Whilst NPC transport factors typically feature multiple 'low-affinity' binding sites for FG domains, these sites remain discrete. Per-site interaction strengths in the lower mM range (*19*, *58*, *59*), equivalent to unbinding free energies of $\sim 5k_{\mathrm{B}}T$, can reduce the diffusivity by an order of magnitude compared to homogeneously attractive colloids (Eq. (20)). The discreteness of interactions thus is a plausible candidate for the reduced transport rates in NPCs.

NPCs feature a variety of nucleoporin FG domains, with the body of available structural and biochemical data suggesting that the cohesiveness of nucleoporin FG domains is highest in the centre and decreases towards the periphery of the pore (*19*, *60*). Qualitatively, one can envisage that the increased solubility of peripheral FG domains promotes a more extended polymer cap, thus minimising total pore resistance and maximising transport rates for strongly attractive colloids. The reduced solubility of the central FG domains, on the other hand, would minimize the size threshold for gating of non-adhesive colloids. Moreover, the accumulation of attractive colloids in the pore may also modulate their transport rates (*61*). These features clearly are not essential for permselective transport through polymer-filled mesopores, but may further enhance selectivity or rates. Our model may be further extended to incorporate solubility gradients and to explore such phenomena in more detail.

### Towards technological applications of synthetic polymer-filled mesopores

Our predictive theoretical approach paves the way for the rational design of nanoporous materials with enhanced selectivity tailored to specific functional requirements, sought after for applications in nanomedicine, biotechnology, and environmental engineering. Mixtures of biological colloids such as folded proteins and other biomacromolecular complexes, as well as synthetic colloids such as nanoparticles, may be effectively separated, not only according to their size but also their surface (bio-)chemistry.

Individual pores, as considered here, are routinely deployed in current nanopore sensing technologies. These technologies enable detection and characterization of individual macromolecules as they travel across the pore (*62*, *63*). Our findings suggest polymer fillings as an attractive tool to develop the next generation of biomolecular sensors with improved sensing capabilities (*64*). Placing a suitable polymer filling upstream the pore's sensing region would enable pre-selection of target solutes from complex mixtures for a focused analysis by the pore. Polymer fillings may also be placed in the very sensing region of the pore to enhance both selectivity and sensitivity.

Individual pores will though typically be insufficient in applications that focus on high-throughput separation (such as filtration systems) delivery (such as porous particles) or catalysis. This limitation can be overcome by multiplexing, e.g., with materials featuring a large array of mesopores. Our theoretical approach remains valid for such arrays as long as the distance between pores remains sufficiently large for the diffusion trajectories of adjacent pores not to substantially interfere. Fortunately, this condition can be met with a relatively tight packing of pores (see iso-concentration lines in Figure 6) (*52*). An avenue not considered here but worthy exploring to increase transport rates further is a pressure gradient that drives solution flow across the mesoporous membrane.

The main design concepts emerging from our theory are:

**1.** To provide selective transport, high permeation selectivity must be coupled with low resistance to diffusive flux. We refer to this combination as 'gating' behaviour, where a minor change in colloid size (Figure 7) or polymer-colloid interaction strength (Figure 5) dramatically shifts the permeation rate from facilitated transport to virtually complete blockage. The thresholds for gating can be tuned by the polymer-solvent interaction strength, and gating is more pronounced for larger colloids.

**2.** The maximal permeability is limited by the resistance of the exterior region. Whilst strong polymer-colloid attraction can make the resistance of the pore interior effectively vanish ($R_{\mathrm{int}} \to 0$), mass transport in plain solvent always provides a non-vanishing resistance of the exterior. Polymer fringes of radius $r_{\mathrm{ext}}$ at the pore entrance and exit decrease the path through plain solvent, and can reduce the external resistance by a factor of up to $\frac{\pi r_{\mathrm{ext}}}{2 r_{\mathrm{p}}}$.

**3.** Pore resistance is highly sensitive to parameters that influence the insertion free energy. The osmotic contribution to the insertion free energy scales as $d^3$ while the interfacial contribution comprises $\chi_{\mathrm{PC}}$ and scales as $d^2$. Thus, a slight change in $d$ and/or $\chi_{\mathrm{PC}}$ translates into a drastic change in permeability.

**4.** A homogeneous polymer-colloid interaction is preferable over one or multiple distinct binding sites to maximise the diffusivity of sticky colloids in the polymer phase, and thus the overall transport rate.

The manufacturing of functional mesoporous membranes is an emerging art (*3*, *4*), and we hope that our theoretical efforts will both promote and guide future practical developments in this area.

## Acknowledgments

The work was supported by the Russian Science Foundation (grant 23-13-00174). M.Y.L. performed computer simulations by SF-SCF method and acknowledges support from Agence Nationale de la Recherche (ANR), France, and Deutsche Forschungsgemeinschaft (DFG), Germany, within grant ANR-20-CE92-0044. R.P.R. conceived the study, made comparison of theoretical predictions to experimental data and acknowledges financial support by the Royal Society (grant IEC/R2/202035) and the UK Biotechnology and Biological Sciences Research Council (grant BB/X00158X/1). R.P.R. thanks Charley Schaefer and Paolo Actis (both University of Leeds) for helpful discussions.

**Code availability:** SF-SCF simulations were performed using the software *Namics* (*65*). Custom code for other modeling and data analysis is available at `https://github.com/miklakt/diffusion_polymer_mesopores`.

# Supplementary Information

# How a polymer filling enhances the rate and selectivity of colloid permeation across mesopores

Mikhail Y. Laktionov, Frans A. M. Leermakers, Ralf P. Richter*, Leonid I. Klushin*, Oleg V. Borisov*

*Corresponding authors: Ralf P. Richter (r.richter@leeds.ac.uk);
Leonid I. Klushin (leo@aub.edu.lb); Oleg V. Borisov (oleg.borisov@univ-pau.fr)

## Supplementary Note 1. Estimates of nucleopore geometry and polymer filling

**Nuclear pore geometry.** According to recent structural work (*14*, *15*), the diameter, and to a lesser extent the height, of the inner FG-domain-filled channel of nuclear complexes can vary depending on the cellular environment and species. Diameters between 40 and 80 nm, and heights between 30 and 70 nm, were reported. In our model, we took diameter and height to extend over 56 Kuhn segments, corresponding to 42.6 nm. This choice reflects that the plain cylinder shape with evenly distributed anchors for homopolymers in our model is a simplifcation of the more complex shape, and the heterogeneity in anchorage and composition of FG domains, in the NPC.

**Polymer filling.** To estimate the amount of intrinsically disordered protein material in the NPC, we considered the copy number, and the size of the region predicted to be intrinsically disordered, for each FG nucleoporin in *S. cerevisiae* yeast NPCs. For the copy numbers, we used the data by Kim et al. (*18*), who assessed all yeast nucleoporins except Nup2 through quantitative mass spectrometry. A large subset of nucleoporins (though missing some major FG nucleoporins) was independently analysed by Rajoo et al. (*17*) through quantitative image analysis, with consistent results. The size of the instrinsically disordered regions had been predicted by Yamada et al. (*66*). We included here the total mass of all intrinsically disordered regions irrespective of their FG motif content, as they contribute to the filling of the pore. The total number of FG nucleoporins (excluding Nup2 as non-essential) per NPC thus is 216, and the total content in intrinsically disordered parts amounts to $11.7 \times 10^4$ amino acids (Supplementary Table S3).

In our model, we took a grafting density of $\sigma = 0.02$ and a degree of polymerization of $N = 300$. With the chosen pore geometry, a Kuhn segment length of $a = 0.76$ nm, and 2 amino acids per Kuhn segment, this corresponds to 197 polymers and a total of $11.8 \times 10^4$ amino acids, in close agreement with the properties of the yeast NPC. The total amount of polymer is also in close agreement with an estimated 14.5 MDa FG domain mass in human NPCs (*60*), which corresponds to $13.1 \times 10^4$ amino acids, considering an average mass per amino acid of 110 Da.

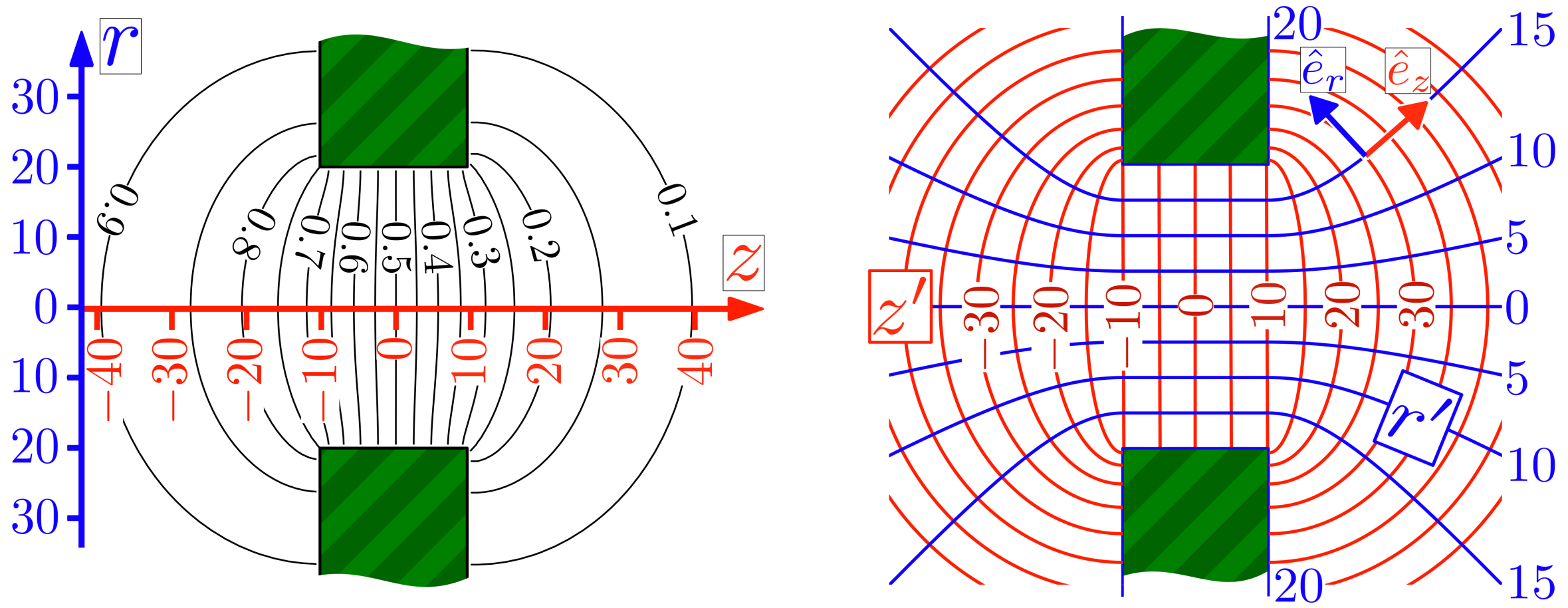

**Figure S1**: **Steady-state diffusion solution and intrinsic coordinate system for a cylindrical pore. Left:** Steady-state solution of the diffusion equation for a point-like particle diffusing through an empty cylindrical pore of finite thickness. Iso-concentration surfaces, $c$ = const, are represented by contour lines with labeled concentration values. Blue and red axes indicate radial and axial coordinates, respectively. **Right:** Intrinsic orthogonal curvilinear coordinate system for the pore. Radial and axial coordinates are parameterized as $r'(r, z)$ and $z'(r, z)$, respectively. Solid lines indicate surfaces of rotation about the pore axis. Red lines correspond to surfaces of constant $z'$; blue lines correspond to surfaces of constant $r'$. Semi-planes with constant angular coordinate $\theta$ are not shown. Local basis vectors of the intrinsic coordinate system ($\hat{e}_r$, $\hat{e}_z$) are illustrated by arrows. Lamé coefficients are defined by the magnitudes of the local basis vectors as $h_r = |\hat{e}_r|$, $h_z = |\hat{e}_z|$, and $h_\theta = |\hat{e}_\theta|$. Pore radius $r_\mathrm{p}^0 = 20$ and thickness $L_0 = 20$ were chosen for illustrative purposes (i.e., to render all numbers clearer than what would be the case with the parameters in Figure 1). The membrane is illustrated in striped green.

## Supplementary Note 2. Analytical estimation of the pore resistance, separating internal and external contributions

**Development of the analytical model in continuous space.** To construct an approximate analytical solution, we first consider a bare pore with a set geometry (Figure 1). The diffusivity $D_0$ is constant throughout the solution. Boundary conditions are set as $c(z = -\infty) = 1$ and $c(z = +\infty) = 0$. The steady-state solution of the diffusion equation, defined by $\partial c/\partial t = 0$, produces the concentration profile shown in Figure S1 (left). For a pore in an infinitely thin membrane, the iso-concentration surfaces are known to form oblate spheroids with the pore rim as their focal circle (*33*). This approximation remains valid for the exterior of the pore in a membrane of finite thickness, as considered here. The solution in the pore lumen is approximated by equally spaced, disk-shaped iso-concentration surfaces. The flux density field is directly related to the concentration gradient via Fick's law $\boldsymbol{j} = -D_0 \nabla c$.

A polymer filling within the pore modifies the local diffusion coefficient $D$ and generates a free-energy landscape, resulting in an effective position-dependent diffusion coefficient (i.e., conductivity) $\tilde{D}(r, z) = D\, e^{-\Delta F/k_B T}$, or local resistivity $\rho = \tilde{D}^{-1}$. Since the flux density $\boldsymbol{j}$ is a conservative vector field, we define a scalar potential function $\psi = c\, e^{\Delta F/k_B T}$, such that $\boldsymbol{j} = -\tilde{D} \nabla \psi$. Consequently, the steady-state iso-concentration surfaces differ from the oblate spheroids of the bare pore; however, they retain a similar structure for iso-values of $\psi$ (Figure 6).

For the exterior region, we introduce an intrinsic curvilinear coordinate system $(r', z', \theta)$ aligned with the approximate iso-surfaces of $\psi$, as depicted in Figure S1 (right) and defined as follows. Level sets of the potential function $\psi$ form a family of oblate hemispheroids $z'$, indexed by their intersection points with the $z$-axis:

$$r' = \{(r, z) \mid \nabla f \cdot \nabla \psi = 0,\ f = f(r, 0)\} \tag{S1}$$

The flux-density stream surfaces, perpendicular to $\psi$, form a family of hyperboloids of revolution $r'$, indexed by their intersection radii with the plane $z = 0$:

$$z' = \{(r, z) \mid \psi(r, z) = \psi(0, z)\} \tag{S2}$$

Half-planes of constant azimuthal angle $\theta$ remain unchanged from the original cylindrical coordinate system.

In the exterior region ($|z| > L/2$, where we consider the effective pore size to account for the finite colloid size), the intrinsic coordinate system $(r', z', \theta)$ is parameterized using the original cylindrical coordinates $(r, z, \theta)$ and the distance from the pore opening $z_{\text{ext}} = |z| - L/2$:

$$r'(r, z) = r\sqrt{1 + \frac{z_{\text{ext}}^2}{r_{\text{p}}^2}}, \tag{S3}$$

$$z'(r, z) = z_{\text{ext}}\frac{\sqrt{r_{\text{p}}^2 - r^2}}{r_{\text{p}}} + \text{sign}(z)\frac{L}{2}. \tag{S4}$$

The corresponding Lamé coefficients are:

$$h_r = \frac{\sqrt{r_{\text{p}}^2 + z_{\text{ext}}^2 - r^2}}{\sqrt{r_{\text{p}}^2 - r^2}}, \tag{S5}$$

$$h_z = \frac{\sqrt{r_{\text{p}}^2 + z_{\text{ext}}^2 - r^2}}{\sqrt{r_{\text{p}}^2 + z_{\text{ext}}^2}}, \tag{S6}$$

$$h_\theta = \frac{r\sqrt{r_{\text{p}}^2 + z_{\text{ext}}^2}}{r_{\text{p}}}, \tag{S7}$$

$$\tilde{h}(r, z) = h_r h_\theta h_z^{-1} = \frac{r}{r_{\text{p}}}\frac{r_{\text{p}}^2 + z_{\text{ext}}^2}{\sqrt{r_{\text{p}}^2 - r^2}}. \tag{S8}$$

The conductivity integrated over the oblate hemispheroids in the exterior region is given by:

$$\varrho_{\text{ext}}^{-1}(z) = 2\pi \int_0^{r_{\text{p}}} \tilde{D}\left(r'(r, z), z'(r, z)\right) \tilde{h}(r, z)\, dr. \tag{S9}$$

Inside the pore ($|z| \leq L/2$), assuming no significant radial flux, the conductivity of a disk cross-section at position $z$ is approximated as:

$$\varrho_{\text{int}}^{-1}(z) = 2\pi \int_0^{r_{\text{p}}} \tilde{D}(r, z)\, r\, dr. \tag{S10}$$

Therefore, the total resistances of the exterior and interior regions are respectively obtained by:

$$R_{\text{ext}} = 2\int_{+L/2}^{+\infty} \varrho_{\text{ext}}(z)\, dz, \tag{S11}$$

$$R_{\text{int}} = \int_{-L/2}^{+L/2} \varrho_{\text{int}}(z)\, dz. \tag{S12}$$

As a control, we revisit the resistance of a bare pore of finite thickness to the diffusion of a point-like colloid using Eqs. (S11, S12). For a bare pore, the effective diffusion coefficient is simply $\tilde{D} = D_0$:

$$\varrho^0_{\mathrm{int}}(z) = \frac{1}{D_0 \pi r_{\mathrm{p}}^2}, \tag{S13}$$

$$\varrho^0_{\mathrm{ext}}(z) = \frac{1}{2\pi D_0 \left(z_{\mathrm{ext}}^2 + r_{\mathrm{p}}^2\right)}. \tag{S14}$$

Integration over the full domains of $z$ yields the expected classical result (*34*) stated in Eq. (9):

$$R^0_{\mathrm{ext}} = 2\int_{-\infty}^{-L/2} \varrho^0_{\mathrm{ext}}(z)\, dz = \frac{1}{2D_0 r_{\mathrm{p}}}, \tag{S15}$$

$$R^0_{\mathrm{int}} = \int_{-L/2}^{+L/2} \varrho^0_{\mathrm{int}}(z)\, dz = \frac{L}{D_0 \pi r_{\mathrm{p}}^2}. \tag{S16}$$

**Application to a discrete cylindrical lattice.** Our numerical approach inherently employs the discrete cylindrical lattice from the SF-SCF calculations (Supplementary Note 3), with discretization steps $\delta z = \delta r = 1$. Consequently, continuous integration across equipotential surfaces foliating space is replaced by summation over discrete conductive layers, each bounded by two adjacent equipotential surfaces. Layers are indexed by the first equipotential surface intersecting the $z$-axis (see Figure S2).

In the interior region, this discretization is straightforwardly applied, giving the resistance of a discrete disk-shaped layer as:

$$\varrho^{\mathrm{lat}}_{\mathrm{int}}(z) = \left[\pi \sum_{r=0}^{r_{\mathrm{p}}-1} (2r+1)\, \tilde{D}[r,z]\right]^{-1}. \tag{S17}$$

For numerical integration in the exterior region, we approximate the equipotential surfaces as cylindrical rather than oblate hemispheroidal layers (Figure S2). Thus, instead of nested oblate hemispheroids (or "bowls"), the discretization yields nested cylinders (or "buckets") with increasing radius $r_{\mathrm{ext}} = r_{\mathrm{p}} + |z| - L/2$ and height $z_{\mathrm{ext}} = |z| - L/2$.

The resistance of each cylindrical layer naturally underestimates the corresponding oblate hemispheroidal layer at the same position $z$. To compensate, we introduce a correction factor $p(z)$, which equals the ratio of the resistances of an oblate spheroidal layer to a cylindrical layer with

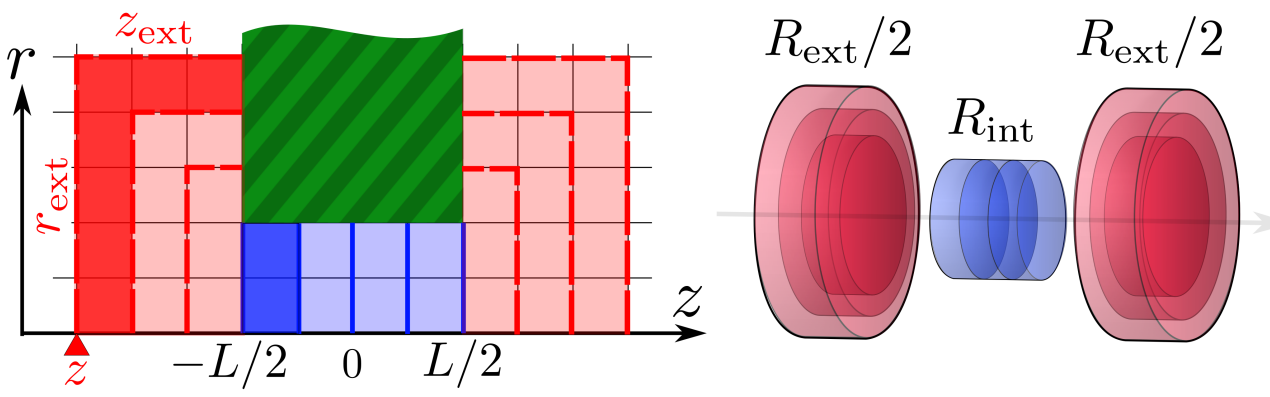

**Figure S2**: **Schematic of the numerical integration of the local conductivity/resistance on a cylindrical lattice. Left:** In the exterior region, conductivities are integrated over nested cylindrical layers (red), each indexed by $z$. A selected layer with radius $r_{ext}$ and height $z_{ext}$ is highlighted with a darker red color. Dashed red lines illustrate the boundaries of the cylindrical layers that approximate the oblate hemispheroidal equipotentials shown in Figure S1. In the interior region, integration is performed over disk-shaped layers (blue), with a selected layer highlighted in darker blue. The membrane is shown in striped green. **Right:** Perspective view to illustrate the nested cylindrical exterior layers (in different tones of red) and disk-shaped interior layers (in different tones of blue) in three dimensions.

identical local conductivity $\tilde{D}$:

$$p(z) = r_p \left( z_{ext} \log \frac{r_{ext}}{r_{ext}-1} + r_{ext}^2 \right)^{-1} \times \\ \times \left[ \operatorname{atan} \frac{z_{ext}}{r_p} - \operatorname{atan} \frac{z_{ext}-1}{r_p} \right]^{-1} . \tag{S18}$$

The resistance of a layer at position $z$ in the exterior region, computed on a cylindrical lattice, then is:

$$\varrho_{ext}^{lat}(z) = \frac{p(z)}{\pi} \left[ \sum_{r=0}^{r_{ext}-1} (2r+1) \tilde{D}[r, z_a] + \right. \\ \left. + 2 \log \left( \frac{r_{ext}}{r_{ext}-1} \right) \sum_{z'=z_a}^{z_b} \tilde{D}[r_{ext}-1, z'] \right]^{-1} \tag{S19}$$

with indexing limits defined by:

$$\begin{cases} z_a = -z, \quad z_b = -L/2 - 1, \quad \text{if } z < -L/2, \\ z_a = L/2, \quad z_b = z - 1, \quad \text{if } z > L/2. \end{cases}$$

To account for the resistance of the semi-infinite reservoirs beyond the integration domain on the cylindrical lattice, we evaluate the analytical expression for the exterior resistance, Equation (S9), from the integration boundary $z$ to infinity:

$$R_{(z,\pm\infty)} = \pm\int_{z}^{\pm\infty} \varrho(z')\,\mathrm{d}z' = \frac{\arctan\left(r_\mathrm{p}/z_\mathrm{ext}\right)}{2D_0\pi r_\mathrm{p}^0} \tag{S20}$$

Finally, the total resistance of the pore, integrated on the discrete cylindrical lattice, is given by:

$$R_\mathrm{int}^\mathrm{lat} = \sum_{z=-L/2}^{+L/2} \varrho_\mathrm{int}^\mathrm{lat}(z) \tag{S21}$$

$$R_\mathrm{ext}^\mathrm{lat} = R_{(z_\mathrm{left},-\infty)} + \sum_{\substack{z\in[z_\mathrm{left},-L/2) \\ z\in(L/2,z_\mathrm{right}]}} \varrho_z + R_{(z_\mathrm{right},+\infty)} \tag{S22}$$

$$R_\mathrm{lat} = R_\mathrm{ext}^\mathrm{lat} + R_\mathrm{int}^\mathrm{lat} \tag{S23}$$

where $z_\mathrm{left}$ and $z_\mathrm{right}$ are the limits of the discrete domain.

Equation (S23) concludes the analytical estimation scheme, giving the total resistance $R_\mathrm{lat}$ of the pore.

## Supplementary Note 3. Computing polymer density maps with the numerical Scheutjens-Fleer self-consistent field (SF-SCF) method

The numerical SF-SCF method was deployed with pore, polymer brush and colloid parameters as defined in Figure S3. With this method, we (i) calculated the spatial polymer distribution $\phi(r, z)$ and (ii) evaluated the insertion free energy for cylindrical colloids into the brush. The latter data served as a reference for the approximate analytical model.

**Main features of the SF-SCF method.** The method is based on the minimization of the excess Helmholtz energy in the incompressible limit, that is, at each specified coordinate $(r, z)$ all volume fractions add up to unity. The method uses lattice approximations, a (local) mean-field approximation with short-ranged interactions parameterized by Flory-Huggins interaction parameters, and employs the freely jointed chain (FJC) model for conformational degrees of freedom of polymer chains composed of segments that fit on the lattice sites.

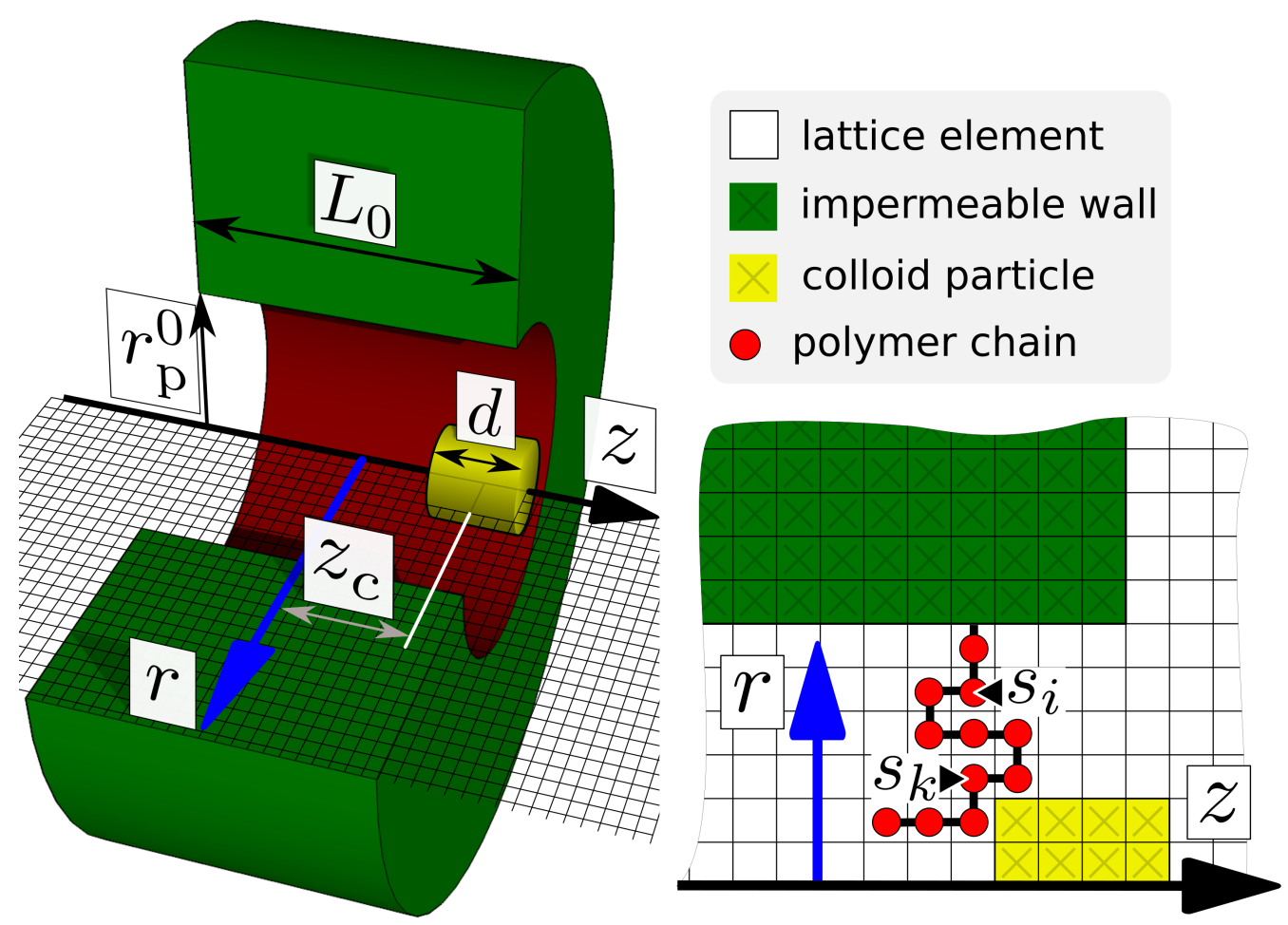

**Figure S3**: **Pore model in the SF-SCF method. Left:** Three-dimensional schematic of the lattice and geometrical features of the cylindrical pore model in the SF-SCF method. **Right:** Key with color code of the modeled objects (top) and their representation on a discrete two-gradient lattice (bottom).

**Discretization and geometrical features of the pore.** As our problem exhibits axial symmetry, space was discretized into a cylindrical lattice with a degenerate angular direction, implemented as a homogeneously curved two-gradient lattice defined by longitudinal $z$ and radial $r$ coordinates (Figure S3). This $(r, z)$ coordinate system is visualized as a two-dimensional Cartesian coordinate system (Figure S3, right); however, each element of the lattice represents a square toroid (of volume $2\pi r a^3$) instead of a square (of area $a^2$). The mean-field approximation is applied in the angular direction, meaning properties in the angular direction are uniform.

The shape of the membrane and any colloids were fixed. The membrane was modeled as a toroid with a rectangular cross-section of physical height $L_0$, a physical inner radius $r_\mathrm{p}^0$, and an outer radius large enough to be effectively infinite with respect to the polymer distribution (Figure S3, green). For $L_0/2 < z < L_0/2$, we refer to the coordinates $\{z_\mathrm{p}\}$ inside the pore. The colloid was modeled as a cylinder with height and diameter $d$ (Figure S3, yellow). The membrane and colloid lattice elements were modeled as impermeable to the solvent and the polymers, illustrated as crossed-out cells in matching colors in Figure S3 (right).

Each polymer chain was represented as a FJC with $s = 1, \cdots, N$ segments of length $a$, modeled as a step-weighted random walk on the lattice (Figure S3, right; red circles with black connecting lines). The first segment $s = 1$ is constrained to the coordinate next to the pore wall (grafting condition). The weights of each step direction were set according to the boundary conditions and the lattice curvature. Steps into impermeable lattice elements had zero weight. Other steps were weighted according to the change in volume per lattice element, with steps towards increasing $r$ consequently being favored, and steps towards decreasing $r$ being disfavored, compared to steps along the $z$-direction. The resulting local polymer concentration $\phi$ is a weighted sum of all possible paths the chain can take (Eq. (S27)).

**Minimization of excess Helmholtz energy.** The excess Helmholtz energy was minimized through a Lagrangian with multipliers $\alpha(r, z)$:

$$\begin{aligned} &F[\boldsymbol{u}, \boldsymbol{\phi}, \boldsymbol{\alpha}] = \\ &= F_{\text{mix}}[\boldsymbol{u}] - \sum_{r,z} \sum_{X} u_X(r, z)\phi_X(r, z) + \\ &+ F_{\text{int}}[\boldsymbol{\phi}] + \sum_{r,z} \alpha(r, z) \left( \sum_{X} \phi_X(r, z) - 1 \right), \end{aligned} \tag{S24}$$

where $\phi_X(r, z)$ is the local volume concentration function of segment type $X$ (polymer, colloid, or solvent), $u_X(r, z)$ is the potential field of segment type $X$, the functional $F_{\text{mix}}[\boldsymbol{u}]$ is the mixing free energy expressed in segment potentials, and the functional $F_{\text{int}}[\boldsymbol{\phi}]$ is the interaction energy expressed in segment densities. The second therm $-u \cdot \phi$ transforms the free energy from potential space to density space.

The condition for the minimum of the functional is a system of three variational equations:

$$\begin{cases} \frac{\partial F}{\partial \boldsymbol{\alpha}} = 0 \\ \frac{\partial F}{\partial \boldsymbol{\phi}} = 0 \\ \frac{\partial F}{\partial \boldsymbol{u}} = 0 \end{cases} \tag{S25}$$

The first condition in Eq. (S25) ensures system incompressibility for each specified coordinate $(r, z)$.

The second condition in Eq. (S25) results in the segment potential field equation for a regular solution:

$$u_A(r, z) = \sum_{B} \chi_{A,B} \left( \langle \phi_B(r, z) \rangle - \phi_B^b \right) + \alpha(r, z), \tag{S26}$$

where $\chi_{A,B}$ is the Flory interaction parameter between segments $A$ and $B$, and $\phi_B^b$ is the volume fraction of $B$ in the bulk (equal to 1 for the solvent and zero otherwise). The angular brackets indicate the site average $\langle X(r,z)\rangle = \sum_{r'=r-1,\ r,\ r+1} \sum_{z'=z-1,\ z,\ z+1} \lambda(r',r,z',z)X(r',z')$ with a priori step probabilities $\lambda(r',r,z',z) = \lambda_{r'-r}(r)\lambda_{z'-z}(z)$ and individual steps obeying to $\lambda_{-1}(r) = \frac{1}{6}\left(1 - \frac{1}{2r-1}\right)$, $\lambda_1(r) = \frac{1}{6}\left(1 + \frac{1}{2r-1}\right)$, $\lambda_{-1}(z) = \lambda_1(z) = 1/6$ and $\lambda_0(z) = \lambda_0(r) = 4/6$. It is easily seen that the sum over the step probabilities equals unity (as it should): $\sum_z' \sum_r' \lambda(r',r,z',z) = 1$.

The third condition in Eq. (S25), the minimization with respect to potentials, links the chain partition function with the local polymer concentration $\phi$ in a diffusion-like equation (Eq. (S32)).

Any subchain of the FJC can be considered a Markov process starting at some segment $s_i$ at coordinates $r_i, z_i$ that goes through intermediate steps to segment $s_k$ at coordinates $r_k, z_k$ (Figure S3, bottom right). Such a process has a statistical weight $G(\{r_k, z_k\}, s_k|\{r_i, z_i\}, s_i)$.

All the Markov processes that start with segment $s_i$ and end with segment $s_k$ at fixed coordinates $\{r, z\}$ are found as the sum over all possible and allowed starting coordinates:

$$G(\{r,z\}, s_k|s_i) = \sum_{r',z'} G(\{r,z\}, s_i|\{r',z'\}, s_i) \tag{S27}$$

The statistical weight of all possible processes that start from segment $s_i$ and end with segment $s_k$ is the sum over all possible coordinates:

$$G(s_k|s_i) = \sum_{r,z} G(\{r,z\}, s_k|s_i)\pi(2r-1) \tag{S28}$$

where the weight $\pi(2r-1)$ is the degeneracy in the radial direction. When $s_i = 1$ and $s_k = N$, the result contains the statistical weight of all possible and allowed conformations of the chain and is the single-chain partition function $G(N|1)$.

There are two initial (starting) conditions. The first one, $G(\{r,z\}, 1|1) = G(r,z)\delta_{r_p,z_p}(r,z)$ is the initial condition of the Markov process which contains just one segment (starts and ends at segment 1). Here $\delta_{r_p,z_p}(r,z) = 1$ when $r_p - r = 1$ and $z \in \{z_p\}$ and zero otherwise (implementing the grafting of chains by segment 1 onto the pore wall). The starting condition for the complementary statistical weights is $G(\{r,z\}, N|N) = G(r,z)$, because these ends are not restricted.

The segment potential $\boldsymbol{u}$ acts on the 'free' segment; thus, Boltzmann statistical weights are applied:

$$G(r,z) = \exp(-u(r,z)) \tag{S29}$$

Within the FJC approach, two complementary propagators are defined, which generate the statistical weights once integrated as described above:

$$G(r,z,s|1) = G(r,z), \langle G(r,z,s-1|1) \rangle, \tag{S30}$$

$$G(r,z,s|N) = G(r,z), \langle G(r,z,s+1|N) \rangle, \tag{S31}$$

where the angular brackets denote averaging over all allowed orientations of the next segment, in analogy to the averaging used for the site fractions.

The volume density distribution of segment $s$ at coordinates $\{r,z\}$ is found from the composition law:

$$\phi(\{r,z\},s) = \frac{2\pi r_{\mathrm{p}}^{0}\, L_0\, \sigma}{G(N,1)} \frac{G(\{r,z\},s|1)\; G(\{r,z\},s|N)}{G(r,z)}, \tag{S32}$$

where $\sigma$ is the grafting density.

Finally, the volume concentration at coordinates $r, z$ is found as the sum over all chain segments:

$$\phi(r,z) = \sum_{s=1}^{N} \phi(\{r,z\},s) \tag{S33}$$

For the solvent, $S$, we may use

$$\phi_S(r,z) = \varphi_S^{\mathrm{b}} \exp(-u_S(r,z)) \tag{S34}$$

where $\varphi_S^{\mathrm{b}} = 1$ represents the plain bulk solution far from the polymer brush.

**Numerical algorithm.** The numerical algorithm solves the Scheutjens-Fleer system of nonlinear equations such that the segment potentials $\boldsymbol{u}$ are consistent with the volume concentrations $\boldsymbol{\phi}$. The relationship between the segment potentials and the volume concentrations is defined in Eq. (S26). Hence, the SF-SCF scheme can be summarized as:

$$\boldsymbol{u}[\boldsymbol{\phi}] \overset{\sum_X \phi_X = 1}{\leftrightarrow} \boldsymbol{\phi}[\boldsymbol{u}] \tag{S35}$$

The equations were solved by a Newton/quasi-Newton optimization routine which effectively leads to the minimization of the functional Eq. (S24). During a numerical iteration step the specified guess for the segment potentials $\boldsymbol{u}$ leads to an update of the volume concentrations $\boldsymbol{\phi}$. It is checked whether these new volume fractions obey to the incompressibility limit, and the values of $\alpha(r,z)$ is adjusted if they do not. The segment potentials are then re-evaluated, and a new guess generated when the input and output potentials differ. The evaluation loop is repeated until the desired accuracy is reached, coresponding to a difference between input and output potentials and errors with respect to the incompressibility constraint of less than 7 significant digits. From the resulting SCF solution, the excess free energy is then evaluated.

For the calculations, we used the package *Namics* developed at Wageningen University, and we refer to the literature for further computational details (*67*).

## Supplementary Note 4. Extracting surface and volume contributions to the colloid insertion free energy from SF-SCF data

The position-dependent insertion free energy can be calculated numerically using the two-gradient SF-SCF method, $\Delta F_{\text{SF-SCF}}(z_c)$, with the restriction that the cylindrical colloid must be placed coaxially along the $z$-axis, with its center located at $z_c$ (see Figure S3).

In contrast, our analytical approach (Supplementary Note 2) allows to compute the insertion free energy for arbitrarily positioned colloids (Eq. (15)). In this approach, the osmotic contribution can be obtained from the polymer volume fraction using the mean-field Flory expression (Eq. (16)). The interfacial term includes both entropic and enthalpic effects: on one hand, the presence of the colloid surface restricts polymer conformations, creating an entropic penalty; on the other hand, polymer-colloid contacts produce an enthalpic effect governed by the interaction parameter $\chi_{\text{ads}}$. At $\chi_{\text{ads}} = \chi_{\text{ads}}^{\text{crit}}$, these entropic and enthalpic contributions cancel, so that the interfacial term vanishes. Additional effects arise because the colloid perturbs the local polymer distribution, producing zones enriched or depleted in polymer concentration around its surface. To capture these local perturbations we introduce phenomenological parameters $b_0$ and $b_1$ to the analytical model, leading to the expression in Eq. (17).

A natural point of comparison between the two approaches is a cylindrical colloid aligned coaxially along the $z$-axis. In this case, insertion free energies can be obtained by both SF-SCF and the analytical model, enabling direct validation. We use this configuration as a reference to determine $b_0$ and $b_1$ by SF-SCF results. Specifically, we fit $b_0$ and $b_1$ by minimizing

$$\sum \left[\Delta F_{\text{SF-SCF}} - \Delta F_{\text{cyl}}(b_0, b_1)\right]^2, \tag{S36}$$

where $\Delta F_{\text{cyl}}(b_0, b_1)$ is the analytical result for the coaxial cylindrical colloid case. Below, we describe both methods in detail and present the fitted values of $b_0$ and $b_1$.

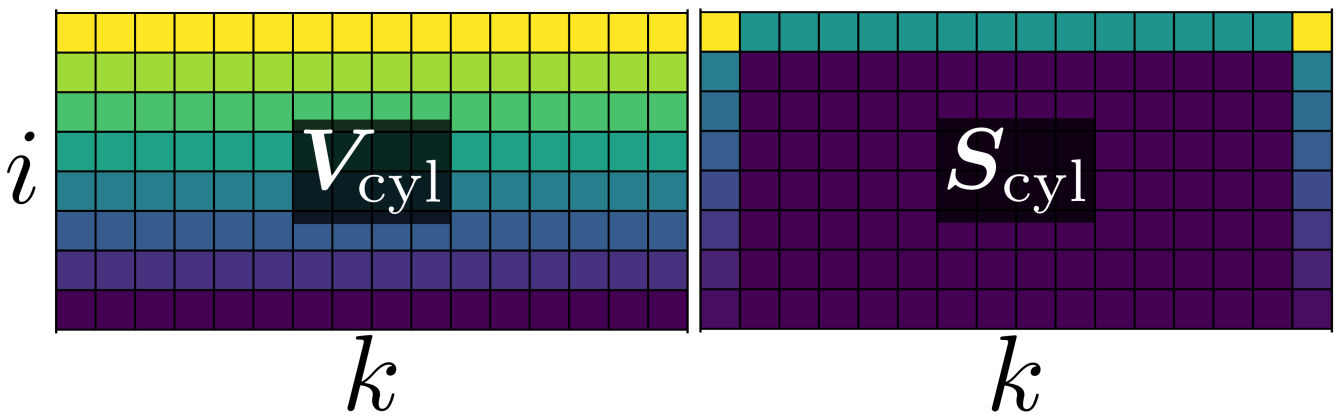

**Figure S4**: **Projection matrices for a cylindrical colloid.** Volume (**left**) and surface (**right**) projection matrices for a cylindrical colloid with diameter and height $d = 16$. The element values of the matrices are color-coded, with violet representing zero and yellow the highest values.

**Analytical estimation of $\Delta F^{\mathrm{osm}}_{\mathrm{cyl}}$ and $\Delta F^{\mathrm{sur}}_{\mathrm{cyl}}$ for cylindrical colloids on a discrete lattice.** Following Eq. (15) the insertion free energy a coaxial cylindrical colloid:

$$\Delta F_{\mathrm{cyl}}(b_0, b_1) = \Delta F^{\mathrm{osm}}_{\mathrm{cyl}} + \Delta F^{\mathrm{sur}}_{\mathrm{cyl}}(b_0, b_1). \tag{S37}$$

where $\Delta F_{\mathrm{cyl}}(b_0, b_1)$ is essentially parametrized by $b_0$ and $b_1$ (see Eqs. (15, 17)).

In their continuous form, the contributions are:

$$\Delta F^{\mathrm{osm}}_{\mathrm{cyl}}(z_{\mathrm{c}}) = 2\pi \int_{z_{\mathrm{c}}-d/2}^{z_{\mathrm{c}}+d/2} \int_0^{d/2} \Pi(r, z)\, r\, dr\, dz \tag{S38}$$

and

$$\begin{aligned} \Delta F^{\mathrm{sur}}_{\mathrm{cyl}}(z_{\mathrm{c}}) = 2\pi d \int_{z_{\mathrm{c}}-d/2}^{z_{\mathrm{c}}+d/2} \gamma(d/2, z)\, dz + \\ +\pi \int_0^{d/2} [\gamma(z_{\mathrm{c}} - d/2, r) + \gamma(z_{\mathrm{c}} + d/2, r)]\, dr, \end{aligned} \tag{S39}$$

where the first term in Eq. (S39) integrates over the lateral surface, and the second term integrates over the top and bottom faces, of the cylinder.

Following the lattice discretization of SF-SCF outputs with discretization steps $\delta r = \delta z = 1$, we use an indexing such that $0 \le i \le d/2 - 1$ iterates in the direction of the $r$-axis, and $0 \le k \le d - 1$ iterates in the direction of the $z$-axis. This corresponds to the physical coordinates $r, z \in [0, d/2] \times [z_{\mathrm{c}} - d/2, z_{\mathrm{c}} + d/2]$ (as illustrated in Figure S3) with $d$ being an even integer to match the lattice.

We define the volume projection matrix $\boldsymbol{V}_{\mathrm{cyl}}[d/2 \times d]$ for a cylindrical colloid of size $d$, such that each element of the matrix equals the volume of the colloid contained within the corresponding

lattice element:

$$V_{\rm cyl}[i,k] = \pi(2i+1) \tag{S40}$$

Obviously, the sum of all matrix elements equals the volume of the cylinder.

$$\sum_{i=0}^{d/2-1} \sum_{k=0}^{d-1} V_{\rm cyl}[i,k] = \frac{\pi d^3}{4}$$

Analogously, we define the colloid surface projection matrix $\boldsymbol{S}_{\rm cyl}[d/2 \times d]$, such that each element of the matrix equals the surface area of the colloid within the corresponding lattice element:

$$\begin{aligned} S_{\rm cyl}[i,k] = &\begin{cases} 2\pi i, & \text{if } i = d/2-1 \\ 0, & \text{otherwise} \end{cases} + \\ &+ \begin{cases} 2\pi(i+1), & \text{if } k=0 \text{ or } k=d-1 \\ 0, & \text{otherwise} \end{cases} \end{aligned} \tag{S41}$$

Here, the first term accounts for the top and bottom faces, and the second term accounts for the lateral surface elements, of the cylinder. Again, the sum of all matrix elements equals the surface area of the cylinder:

$$\sum_{i=0}^{d/2-1} \sum_{k=0}^{d-1} S_{\rm cyl}[i,k] = \frac{3\pi d^2}{2}$$

Figure S4 provides a color-coded map of the volume $\boldsymbol{V}_{\rm cyl}$ and surface $\boldsymbol{S}_{\rm cyl}$ projection matrices for a selected colloid size ($d = 16$).

To calculate the colloid insertion free energy, we integrated the osmotic pressure over the colloid volume and the surface tension over the colloid surface (Eq. (15)). The equivalent operation on a discrete lattice is the matrix dot product. The two contributions to the insertion free energy are thus calculated as:

$$\begin{aligned} \Delta F^{\rm osm}_{\rm cyl}(z_{\rm c}) &= \boldsymbol{V}_{\rm cyl} \cdot \boldsymbol{\Pi}\{z_{\rm c}\} \text{ and} \\ \Delta F^{\rm sur}_{\rm cyl}(z_{\rm c}) &= \boldsymbol{S}_{\rm cyl} \cdot \boldsymbol{\gamma}\{z_{\rm c}\}, \end{aligned} \tag{S42}$$

where the matrix elements for $\boldsymbol{\Pi}\{z_{\rm c}\}$ and $\boldsymbol{\gamma}\{z_{\rm c}\}$ run across $0 \le i < d/2$ and $z_{\rm c} - d/2 \le k < z_{\rm c} + d/2$.

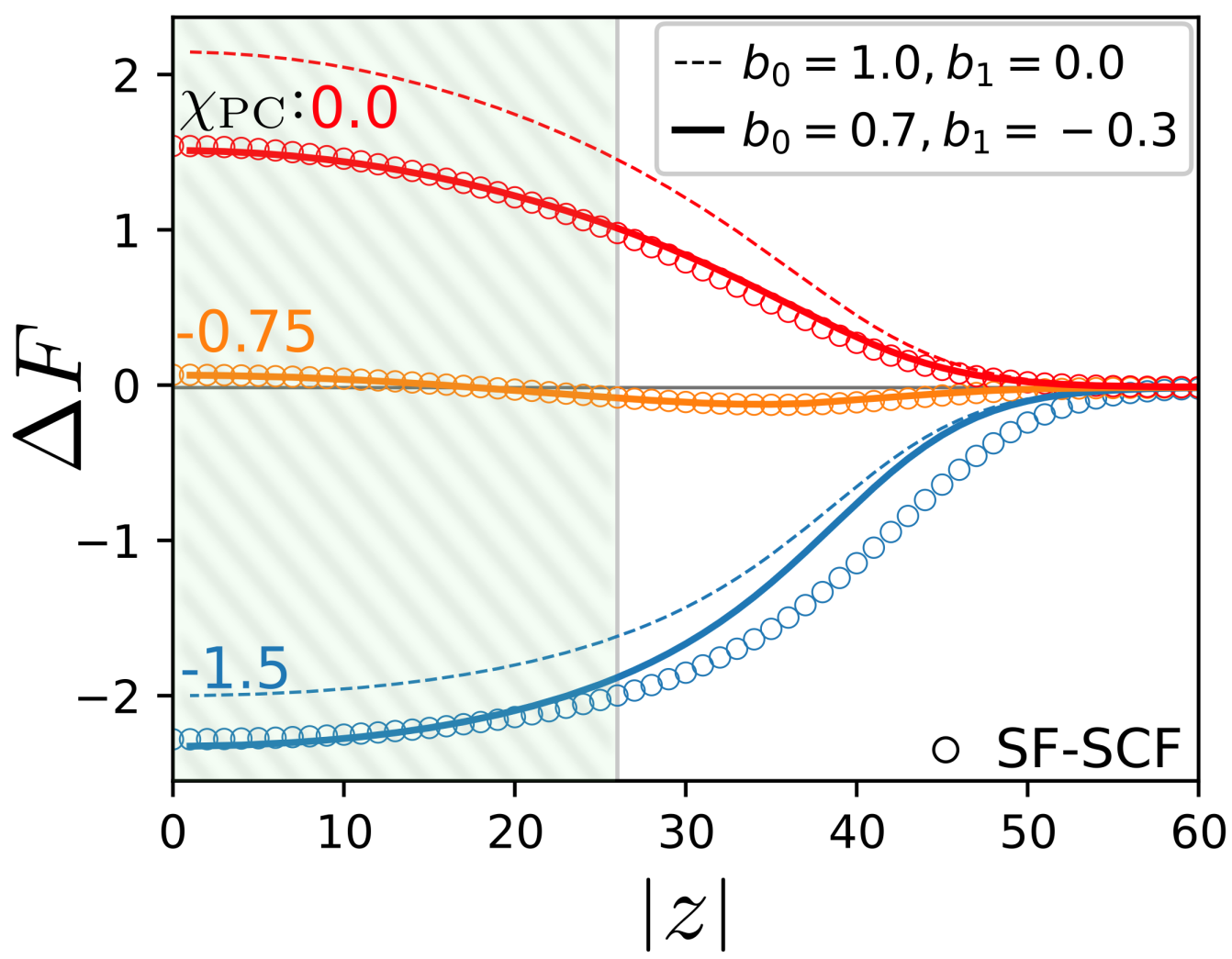

**Figure S5**: **Effect of fitting parameters on the match of analytical model to SF-SCF data.** Comparison of $\Delta F_{\text{SF-SCF}}$ profiles (circles) with $\Delta F_{\text{cyl}}(b_0, b_1)$, for the best-fit values of $b_0 = 0.7$ and $b_1 = -0.3$ (optimally accounting for local perturbations in the polymer concentration; thick solid lines) and for $b_0 = 1.0$ and $b_1 = 0.0$ (neglecting any local perturbations in polymer concentration; thin dashed lines). Pore and brush parameters are as given in Figure 1; $d = 4$, $\chi_{\text{PS}} = 0.5$, and $\chi_{\text{PC}}$ values are color-coded and indicated on the right side of the figure. The light green hatched area marks values of $z_c$ that are located inside the pore lumen ($|z| \leq 26$).

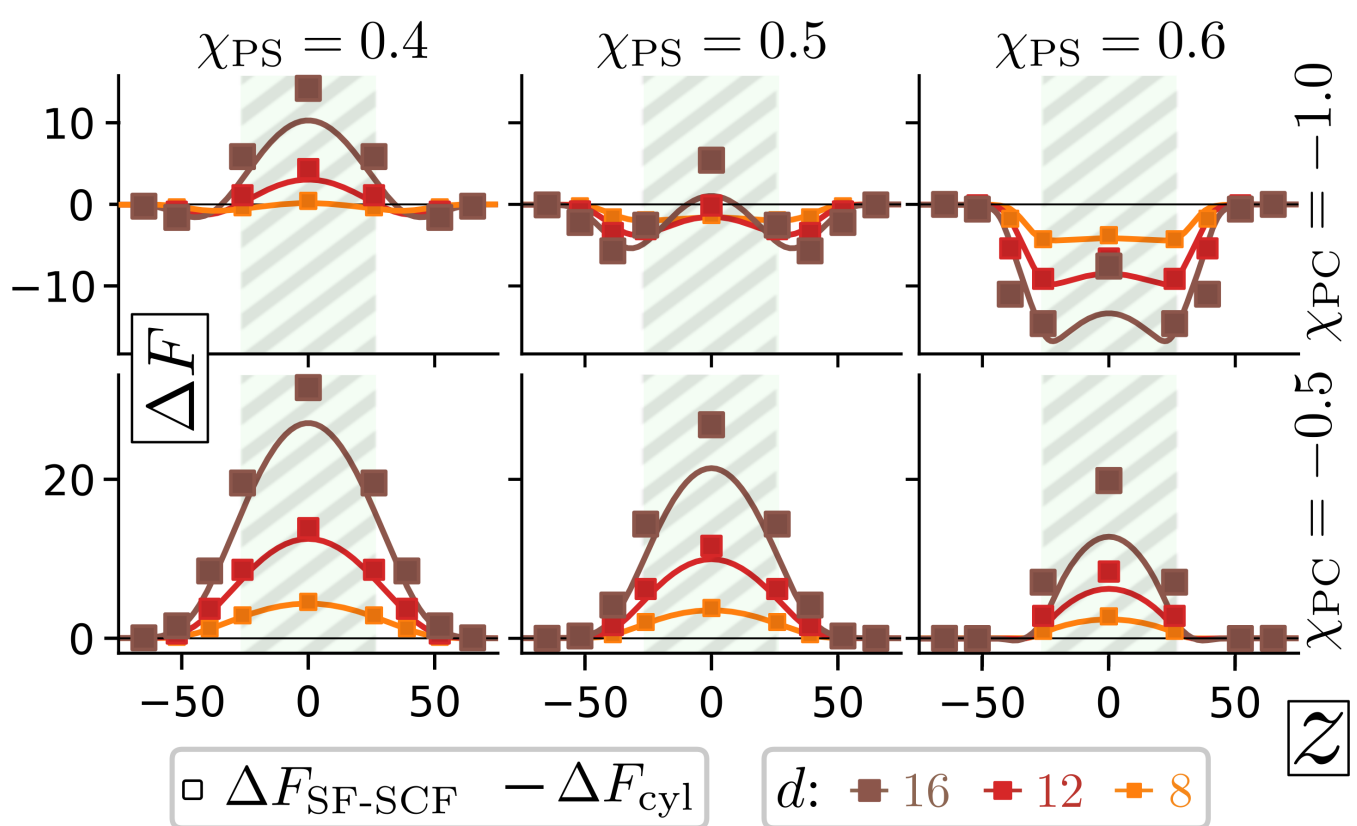

**Figure S6**: **Effect of cylindrical colloid size on the match of analytical model to SF-SCF data.** Comparison of $\Delta F_{\text{SF-SCF}}$ (square symbols) with $\Delta F_{\text{cyl}}(b_0, b_1)$ profiles for the best-fit values $b_0 = 0.7$ and $b_1 = -0.3$ (solid lines), for $d = [8, 12, 16]$ (color coded). Pore and brush parameters are as given in Figure 1; $\chi_{\text{PC}} = -1.0$ (top row) and -0.5 (bottom row), and the solvent quality was varied near the $\theta$-point with $\chi_{\text{PS}} = [0.4, 0.5, 0.6]$, as indicated. The light green hatched area marks values of $z_c$ that are located inside the pore lumen ($|z| \leq 26$).

**Extracting insertion free energy from SF-SCF data.** As illustrated in Figure S3, a cylindrical colloid is positioned coaxially along the $z$-axis as impermeable lattice elements. Consequently, the polymer-brush chains adjust to the available space and the polymer-colloid interaction strength, producing a change in the system's total free energy, $F_{\text{cyl}}(z_c)$.

We performed a series of SF-SCF calculations to obtain free energy profiles $F_{\text{SF-SCF}}(z_c)$ for a range of colloid diameters $d$, polymer-solvent interaction parameters $\chi_{\text{PS}}$, and polymer-colloid interaction parameters $\chi_{\text{PC}}$, using the pore geometry and brush parameters defined in Figure 1. To isolate the insertion free energy, we apply a ground-state correction and define

$$\Delta F_{\text{SF-SCF}}(z_c) = F_{\text{SF-SCF}}(z_c) - F_{\text{SF-SCF}}(z_c^{\text{bulk}}), \tag{S43}$$

where $F_{\text{SF-SCF}}(z_c^{\text{bulk}})$ is the system free energy when the colloid is placed far away from the pore.

**Fitting $b_0$ and $b_1$ to match analytical model to SF-SCF data.** The optimal coefficients $b_0$ and $b_1$ were found using the least-squares method, i.e., minimizing Eq. (S36) across a range of $\chi_{\text{PS}} \in [0, 1]$ and $\chi_{\text{PC}} \in [-2, 0]$ for the relevant colloid positions $|z_c| \leq 60$. Fits were performed with small colloids ($d = 4$), to focus on local effects and avoid added effects that may arise due to global perturbations of the polymer distribution in the pore.

In a $\theta$-solvent (chosen here as a representative case) and for the selected pore geometry and brush parameters (Figure 1), we selected three $\chi_{PC}$ values (0.00, -0.75 and -1.50) to span the regimes of net repulsion, near-critical adsorption, and net attraction, respectively. Figure S5 demonstrates that a satisfactory fit (thick solid line) to the $\Delta F_{\text{SF-SCF}}$ profiles (circles) could be obtained across all three representative datasets with a single parameter set, $b_0 = 0.7$ and $b_1 = -0.3$. The insertion free energies are consistently close to zero for $\chi_{PC} = -0.75$, illustrating that this value is indeed near the critical condition, $\chi_{PC}^{crit} = \chi_{crit} + \chi_{PS}(1 - \phi)$, where polymer attraction and osmotic repulsion just cancel each other. Consequently, the least-square fits considered the regimes of net repulsion and net attraction approximately equally through $\chi_{PC} = 0.00$ and $\chi_{PC} = -1.50$, respectively. In contrast, neglect of local perturbations to the polymer concentration ($b_0 = 1.0$ and $b_1 = 0.0$) reproduces $\Delta F_{\text{SF-SCF}}$ rather poorly for repulsive and attractive colloids (Figure S5, thin dashed line), thus demonstrating the importance of the correction.

Although the fit was performed only for small colloids, the resulting parameters $b_0$ and $b_1$ still successfully account for the local perturbations to the polymer concentration when calculating the insertion free energy of larger colloids. This is illustrated in Figure S6 for two selected $\chi_{PC}$ values (-0.5 and -1.0) and three selected $\chi_{PS}$ values (0.4, 0.5 and 0.6).

Deviations become notable, however, for $d = 16$ under conditions of strong repulsion or attraction. This size regime thus marks the limit of validity of the local perturbation approximation. Instead, attractive or repulsive interactions entail non-local changes that impact the polymer concentration across the entire pore cross-section. As a consequence, the pore walls also influence colloid insertion.

It is notable in Figure S5, that for strongly attractive colloids ($\chi_{PC} = -1.5$) outside the pore ($|z_c| > 26$) the SF-SCF method (blue circles) predicts systematically lower insertion free energies than the analytical approach (blue solid line). Inspection of SF-SCF polymer density maps (not shown) showed that in this regime the polymer brush changes conformation to reach the colloid at a greater distance $|z_c|$ from the pore. This effect, which is not captured in our analytical approach, is mild and slightly increases the region of negative insertion free energy. It is expected to entail a slight reduction in the total resistance to diffusive transport, as the capture of colloids by extending polymer fringes reduces the resistance of the bulk solution which dominates for attractive pores.

## Supplementary Note 5. Computing insertion free energies for arbitrarily placed spherical colloids

The SF-SCF method considered in the previous sections is limited to colloids moving along the main axis of the pore. In reality, colloids may be located off the pore axis. Here, we generalize the analytical approach (Supplementary Note 4) to calculate insertion free energies based on volume and surface contributions (assuming localized perturbations to the polymer concentration) to arbitrarily placed colloids. In doing so, we also change the shape of the colloid, from a cylinder to a simpler sphere.

**Angular integration of colloid volume and surface onto the $rz$-plane in continuous space.** As the polymer and pore geometrical features are uniform in the angular direction, we use $r, z$ cylindrical coordinates with a degenerate angular axis. Any property such as the polymer volume fraction can hence be expressed as a function $f(r, z, \theta) = f(r, z)$.

The distance to the center of spherical body of a colloid is

$$\Delta_c = \sqrt{r^2 + r_c^2 - 2rr_c\cos(\theta) + (z - z_c)^2}, \tag{S44}$$

where $r_c, z_c, \theta_c$ are the position of the center in cylindrical coordinates. Without the loss of generality, we set $\theta_c = 0$.

The function $f(r, z)$ is integrated over the spherical volume as:

$$\begin{aligned} \int_V f dV &= \int_0^{+\infty}\int_{-\infty}^{+\infty} f(r, z) \int_0^{2\pi} H(\Delta_c - d/2) r\mathrm{d}r\mathrm{d}z\mathrm{d}\theta \\ &= \int_0^{+\infty}\int_{-\infty}^{+\infty} f(r, z) V_{\theta\downarrow}(r, z)\mathrm{d}r\mathrm{d}z \end{aligned} \tag{S45}$$

where $r\mathrm{d}r\mathrm{d}z\mathrm{d}\theta$ is the differential volume element in cylindrical coordinates, $H(\Delta_c - d/2)$ is the Heaviside function, that evaluates to 1 inside the sphere. $V_{\theta\downarrow}(r, z)$ is a projection of sphere volume on the $rz$-plane, with $\theta$ being the projecting direction.

Similarly, to compute surface integrals over the spherical colloid, the Dirac delta function $\delta(\Delta_c - d/2)$ is applied to restrict the integration domain to the spherical surface. This allows to express the surface integral of a scalar function $f(r,z)$ as:

$$\int_S f dS = \\ = \int_0^{+\infty}\int_{-\infty}^{+\infty}\int_0^{2\pi} f(r,z)\delta(\Delta_c - d/2)\frac{\Delta_c}{r_c|\sin\theta|} r\,\mathrm{d}r\mathrm{d}z\mathrm{d}\theta = \\ = \int_0^{+\infty}\int_{-\infty}^{+\infty} f(r,z) S_{\theta\downarrow}\,\mathrm{d}r\mathrm{d}z \tag{S46}$$

where the factor $\frac{\Delta_c}{r_c|\sin\theta|} r\, drdz$ corresponds to the surface area element expressed in cylindrical coordinates.

From Eqs. (S45, S46) the volume $V_{\theta\downarrow}$ and surface $S_{\theta\downarrow}$ projections of a spherical body onto the $rz$-plane in cylindrical coordinates are:

$$V_{\theta\downarrow}(r,z,r_c,z_c) = 2\int_0^{\pi} H(\Delta_{\text{center}} - d/2)\, r\,\mathrm{d}\theta \tag{S47}$$

$$S_{\theta\downarrow}(r,z,r_c,z_c) = 2\int_0^{\pi} \delta(\Delta_c - d/2)\frac{\Delta_c}{r_c|\sin\theta|} r\,\mathrm{d}\theta \tag{S48}$$

These expressions describe the angular integration of the volume and surface projected onto the $rz$-plane, effectively reducing the 3D geometry to a two-gradient description suitable for cylindrical symmetry.

**Application to a discretized lattice.** To find the elements of the projection matrices (each representing the volume or surface area of the spherical colloid intersecting a given lattice cell), the angularly projected volume and surface area are dicretized over finite lattice elements. For each grid element indexed by $i,k$, corresponding to the domain $r \in [i, i+\delta r]$ and $z \in [k, k+\delta z]$, the matrix entries are defined as

$$V\{r_c\}[i,k] = \iint_{i,k}^{\substack{i+\delta r\\ k+\delta z}} V_{\theta\downarrow}(r,z,r_c,z_c)\,\mathrm{d}r\mathrm{d}z \tag{S49}$$

$$S\{r_c\}[i,k] = \iint_{i,k}^{\substack{i+\delta r\\ k+\delta z}} S_{\theta\downarrow}(r,z,r_c,z_c)\,\mathrm{d}r\mathrm{d}z \tag{S50}$$

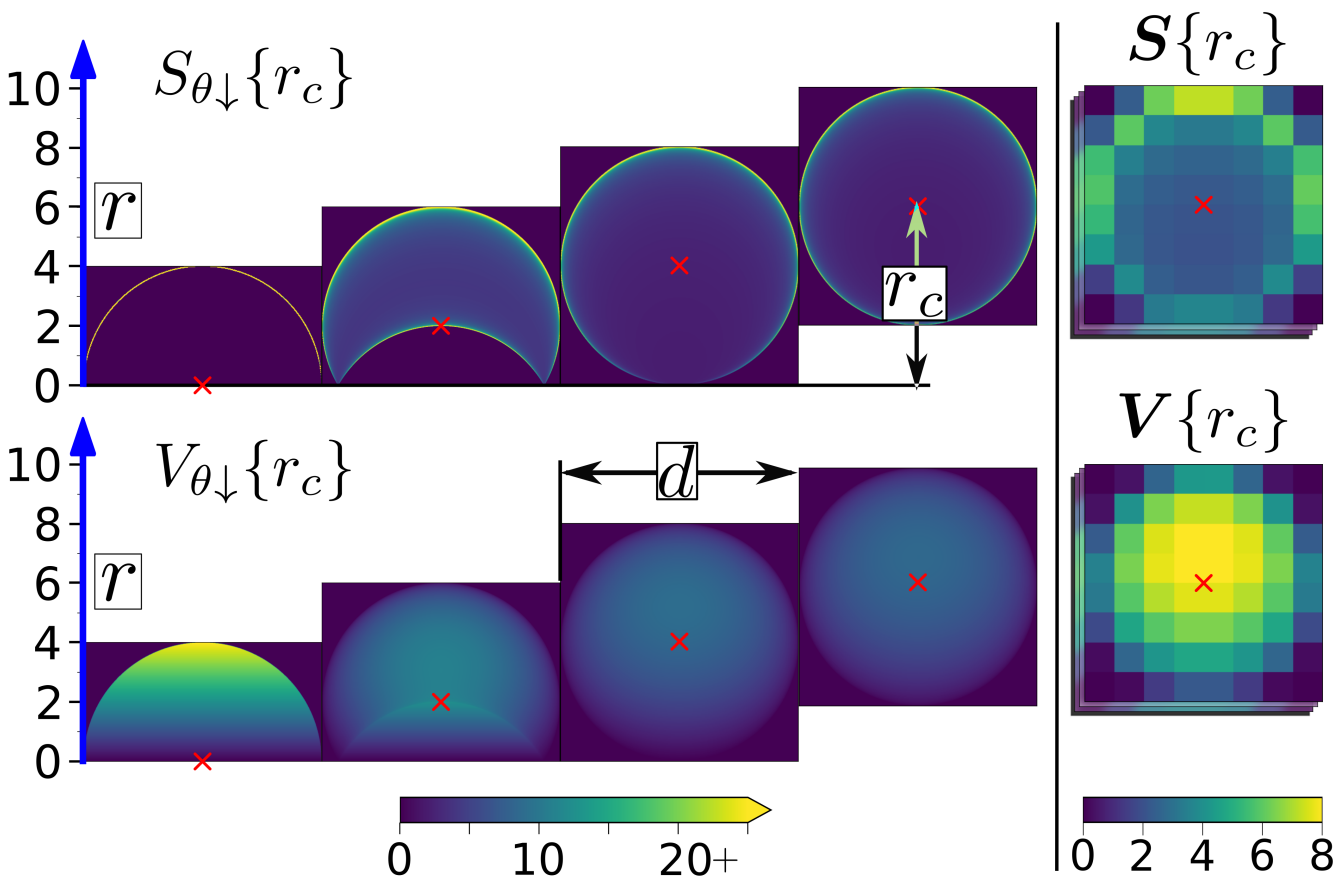

**Figure S7**: **Projections and projections matrices for a sperical colloid. Left:** Surface $S_{\theta\downarrow}\{r_\mathrm{c}\}$ (top) and volume $V_{\theta\downarrow}\{r_\mathrm{c}\}$ (bottom) projections on the $rz$-plane for a spherical colloid with diameter $d = 8$, for a set of sphere center offsets $r_\mathrm{c} = \{0, 2, 4, 6\}$ to the $z$-axis (shown side by side). The sphere centers are indicated with a red cross. The $r$-axis is shown with a blue arrow; the $z$-axis is omitted, as the center coordinates $z_\mathrm{c}$ are arbitrary. **Right:** Discretization results of the surface and volume projection matrices $\boldsymbol{S}\{r_\mathrm{c}\}$ and $\boldsymbol{V}\{r_\mathrm{c}\}$ as heatmaps, for $r_\mathrm{c} = 6$. Color codes are distinct for dsata on the left and right, as indicated. Intensities for surface and volume are normalized by $a^2$ and $a^3$, respectively.

where $z_{\mathrm{c}}$ has an arbitrary value. The size of the projection matrices is $\min(d, r_{\mathrm{c}} + d/2) \times d$. In contrast to Cartesian projections, cylindrical projections depend on the radial coordinate of the center $r_{\mathrm{c}}$.

Naturally, the sum of all matrix elements equals the surface and the volume of the sphere

$$\sum_{\substack{i\in[0,\min(r_{\mathrm{c}}+d/2,d)-1]\\ k\in[0,d-1]}}\sum V\{r_{\mathrm{c}}\}[i,k] = \pi d^2,$$

$$\sum_{\substack{i\in[0,\min(r_{\mathrm{c}}+d/2,d)-1]\\ k\in[0,d-1]}}\sum S\{r_{\mathrm{c}}\}[i,k] = \frac{\pi d^3}{6}.$$

The discretized form of Eq. (15) to calculate the osmotic term in the insertion free energy for a spherical colloid is:

$$\Delta F_{\mathrm{osm}}(r_{\mathrm{c}}, z_{\mathrm{c}}) = \sum_{\substack{i\in[0,\min(r_{\mathrm{c}}+d/2,d)-1]\\ k\in[0,d-1]}}\sum V\{r_{\mathrm{c}}\}[i,k]\ \Pi\left[\max(r_{\mathrm{c}} - d/2, 0) + i,\ z_{\mathrm{c}} - d/2 + k\right], \tag{S51}$$

$$\text{where}\quad \mathbf{\Pi}\{r_{\mathrm{c}}, z_{\mathrm{c}}\} = \left(\Pi_{i,k}\right)_{\substack{\max(r_{\mathrm{c}}-d/2,0)\le i<r_{\mathrm{c}}+d/2\\ z_{\mathrm{c}}-d/2\le k<z_{\mathrm{c}}+d/2}}.$$

Similarly, the surface term in the insertion free energy is the following matrix multiplication:

$$\Delta F_{\mathrm{sur}}(r_{\mathrm{c}}, z_{\mathrm{c}}) = \sum_{\substack{i\in[0,\min(r_{\mathrm{c}}+d/2,d)-1]\\ k\in[0,d-1]}}\sum S\{r_{\mathrm{c}}\}[i,k]\ \gamma\left[\max(r_{\mathrm{c}} - d/2, 0) + i,\ z_{\mathrm{c}} - d/2 + k\right], \tag{S52}$$

$$\text{where}\quad \boldsymbol{\gamma}\{r_{\mathrm{c}}, z_{\mathrm{c}}\} = \left(\gamma_{i,k}\right)_{\substack{\max(r_{\mathrm{c}}-d/2,0)\le i<r_{\mathrm{c}}+d/2\\ z_{\mathrm{c}}-d/2\le k<z_{\mathrm{c}}+d/2}}.$$

The function domain, values and discretization for $V_{\theta\downarrow}$ and $S_{\theta\downarrow}$ are exemplified in Figure S7 for a set of colloids with varying radial center positions $r_{\mathrm{c}}$.

The method was used to compute the insertion free energy $\Delta F(r, z)$ from inherently discrete SF-SCF outputs $\Pi, \gamma$, preserving the cylindrical lattice. Volume and surface projection matrices are explained geometrically in Figure S8.

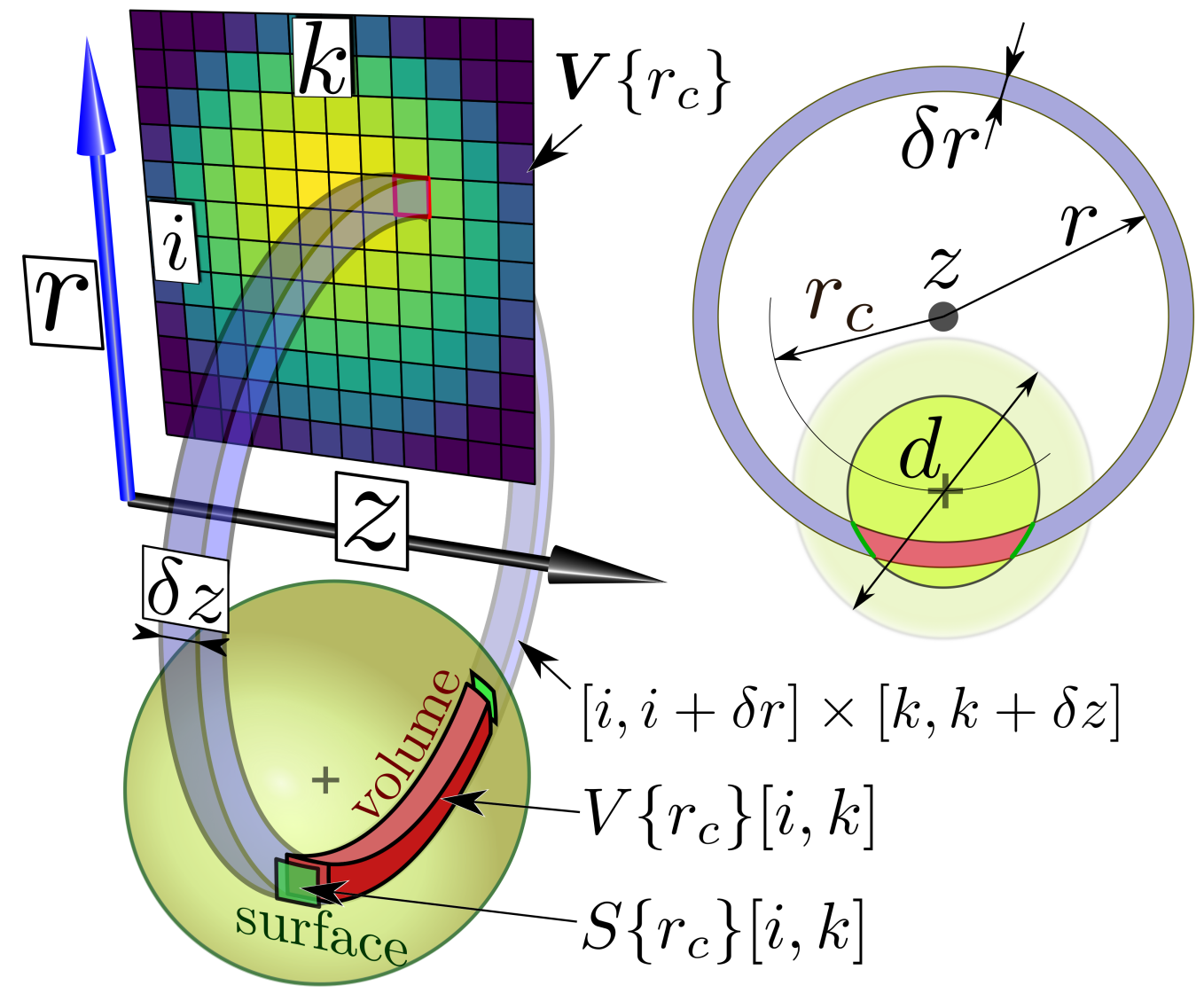

**Figure S8**: **Illustration of a spherical colloid's volume and surface projection matrices in a cylindrical lattice.** Shown on the left is the volume projection matrix $\boldsymbol{V}\{r_c\}$ for a spherical colloid with diameter $d = 12$ and $r_c = 8$. The colored tiles encode the matrix elements' values, where violet means zero and yellow represents the largest values. The geometrical meaning of the matrix element is the colloid volume (red body) or surface (green small patch) found in the domain $r, z \in [i, i + \delta r] \times [k, k + \delta z]$ (opaque blue toroid). The smaller drawing on the right complements the main drawing with a $z$-view for clarity, using consistent color-coding for volume and surface elements, with the pale blue annulus being the domain of lattice element. The yellow circle indicates the colloid cross-section within the selected $z$-slice, while the larger pale yellow circle shows the rest of the colloid body lying behind the plane of the cross-section.

## Supplementary Note 6. Validating analytical approaches with numerical simulations

Applying Eqs. (2, 3) and the boundary conditions reduces the Smoluchowski equation (Eq. (1)) in the stationary state to the generalized Laplace equation:

$$\mathcal{L}\tilde{c} = 0, \tag{S53}$$

with the Smoluchowski diffusion operator

$$\mathcal{L} = \nabla\cdot\left(\tilde{D}(\boldsymbol{r})\nabla\right).$$

**Defining the Laplace equation in a finite, discretized domain.** To obtain a stationary solution of Eq. (S53), we consider a finite cylindrical domain large enough to exclude edge effects:

$$(r, z) \in \Omega = [0, r_{\max}] \times [z_{\min}, 0],$$

and discretize it on a regular cylindrical lattice with spacings $\delta r = \delta z$. Due to symmetry about the mid-plane $z = 0$, we restrict the domain accordingly with suitable boundary conditions.

The computational grid contains $N_r = r_{\max}/\delta r$ radial nodes and $N_z = (z_{\max} - z_{\min})/\delta z$ axial nodes, indexed by $i = 0, \ldots, N_r - 1$ and $k = 0, \ldots, N_z - 1$, respectively. Physical coordinates are $r_i = i\delta r$ and $z_k = z_{\min} + k\delta z$ as shown in Figure S9a. Each node defines a finite volume element with four faces, labeled $z\pm$ and $r\pm$, as illustrated in Figure S9c. The finite volume element is indexed using its lower-left node $(r-, z-)$.

Applying the divergence theorem to each finite volume element yields a discrete approximation to the Laplacian in the form of a finite volume stencil:

$$\begin{aligned}\nabla_{i,k}\psi =& \lambda^z D^{z+}_{i,k}(\psi_{i,k} - \psi_{i,k+1}) + \lambda^z D^{z-}_{i,k}(\psi_{i,k} - \psi_{i,k-1}) + \\ &+ \lambda^{r+}_i D^{r+}_{i,k}(\psi_{i,k} - \psi_{i+1,k}) + \lambda^{r-}_i D^{r-}_{i,k}(\psi_{i,k} - \psi_{i-1,k}),\end{aligned} \tag{S54}$$

where, $D^{r\pm}_{i,k}$ and $D^{z\pm}_{i,k}$ denote effective diffusion coefficient evaluated at finite volume element faces, as a harmonic averaging of values at adjacent finite volume elements. The finite volume expansion in the radial direction (see Figure S9b) and discretization steps are accounted for through

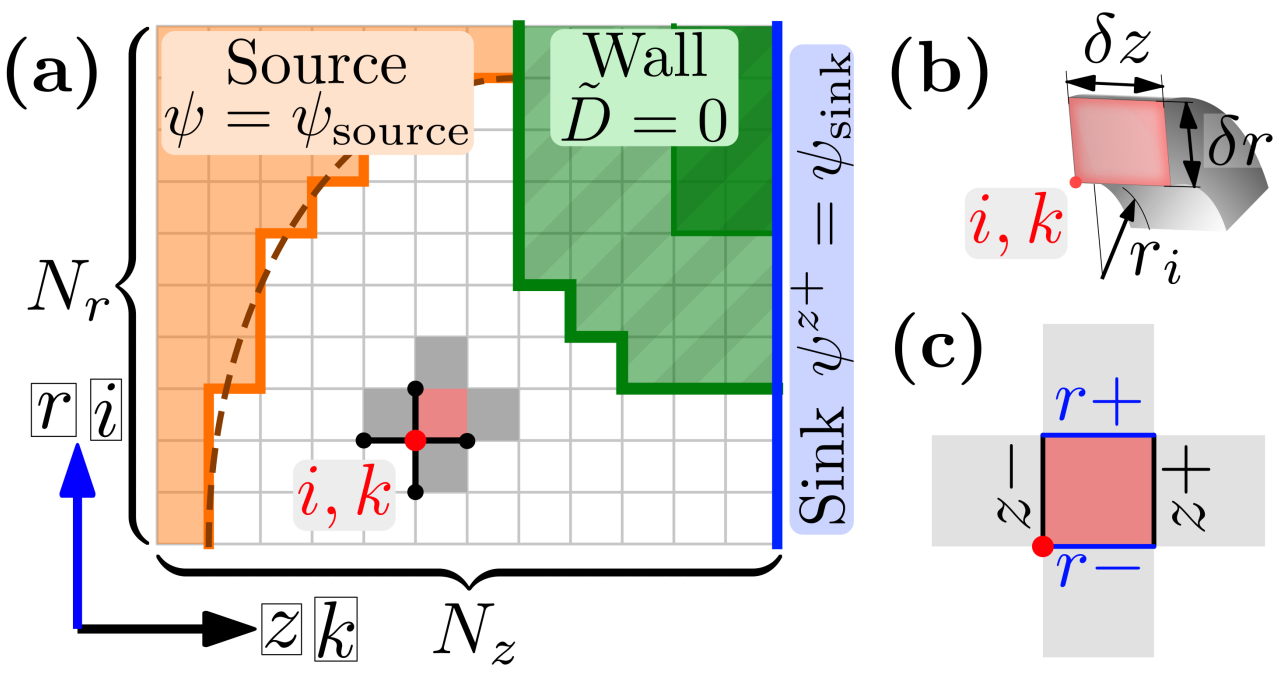

**Figure S9**: **Numerical simulation domain and finite volume elements stencil. (a)** Discretized domain for numerical simulations. Following the analytical solution for the bare pore, the source nodes are placed in the region circumferenced by oblate hemispheroids (orange). The effective shape of the impermeable membrane wall is constructed from the real membrane wall (dark green) and excluded volume due to the finite colloid size (light green, see Figure S10). Exploiting system symmetry, we modelled only half of the system, by placing Dirichlet boundary condition on the mid-plane of the full system. The Smoluchowski operator is defined with a 5-point stencil on the finite volume elements (red element with grey neighbors). The dots on the stencil denote the indexing nodes. **(b)** The finite volume elements naturally expand with the radial coordinate, with the size defined by the discretization steps $\delta r$ and $\delta z$. **(c)** Each volume element shares common faces with the neighboring elements, labeled with the coordinate direction along the radial $r\pm$ and the axial $z\pm$ axes.

coordinate-dependent prefactors:

$$\lambda^z = \frac{1}{\delta z^2}, \tag{S55}$$

$$\lambda_i^{r+} = \frac{2r_i + 2\delta r}{2r_i + 1}\frac{1}{\delta r^2}, \tag{S56}$$

$$\lambda_i^{r-} = \frac{2r_i}{2r_i + 1}\frac{1}{\delta r^2}. \tag{S57}$$

The discretized system is assembled into matrix form:

$$\mathbf{L}\,\boldsymbol{\psi} = \boldsymbol{b}, \tag{S58}$$

where $\mathbf{L}$ applies the finite volume stencil from Eq. (S54) and $\boldsymbol{b}$ enforces boundary conditions.

The two-dimensional grid of volume elements are flattened into vectors of length $N_r N_z$ reindexed with $m = iN_z + k$. Under this mapping, the continuous operator $\mathcal{L}$ becomes a sparse $[N_rN_z \times N_rN_z]$ matrix $\mathbf{L}$, and the unknown values of $\psi(r_i, z_k)$ form the column $[N_rN_z]$ vector $\boldsymbol{\psi}$.

Once Eq. (S58) is solved for $\boldsymbol{\psi}$, the resulting vector is reshaped into the $rz$-grid, giving $\psi(r, z)$ and recovering the colloid concentration $c(r, z)$. This enabled examination of the non-equilibrium partitioning of colloids in the presence of position-dependent insertion free energies for given boundary conditions.

**Defining the impermeable wall (no-flux elements).** The impermeable regions are inherited from the SF-SCF calculations and define the finite volume elements inaccessible to diffusing colloids. We define the original impermeable wall grid elements as:

$$\mathbf{Wall}_0[i,k] = \begin{cases} 1 & \text{if } r_i \geq r_\mathrm{p}^0 \text{ and } z_k \leq -L_0/2 \\ 0 & \text{otherwise} \end{cases} \tag{S59}$$

Due to excluded volume of the colloid, the effective pore shape is different. The space impermeable to the centre of a spherical colloid of diameter $d$ defines an effective pore with radius $r_\mathrm{p} = r_\mathrm{p}^0 - \frac{d}{2}$, length $L = L_0 + d$, and rounded corners, as shown in Figure S10a. In discretized space, the dilated impermeable region is computed via the morphological dilation operation:

$$\mathbf{Wall} = \mathbf{Wall}_0 \oplus \mathbf{Colloid} \tag{S60}$$

with the structuring element **Colloid** being a coarse-grained circle in $rz$-coordinate, see Figure S10b.

$$\mathbf{Colloid}[i,k] = \begin{cases} 1, & \text{if } \left(\frac{d}{2} - r_i\right)^2 + \left(\frac{d}{2} - z_k\right)^2 \leq \frac{d^2}{4} \\ 0, & \text{otherwise} \end{cases} \tag{S61}$$

**Defining the colloid source.** We place source nodes at a finite axial distance, chosen to align with an oblate hemispheroidal surface (or a half ellipse in the $rz$-plane) intersecting the $z$-axis at $z_\mathrm{min} + \delta z$. The half ellipse has foci located at $(\pm r_\mathrm{p}, -L/2)$, such that the pore rim acts as the focal circle. This defines the major semi-axis $r_\mathrm{max} - \delta r$ and hence the radial extent of the discrete computational domain:

$$r_\mathrm{max} - \delta r = \sqrt{(|z_\mathrm{min}| - L/2 + \delta z)^2 + r_\mathrm{p}^2} \tag{S62}$$

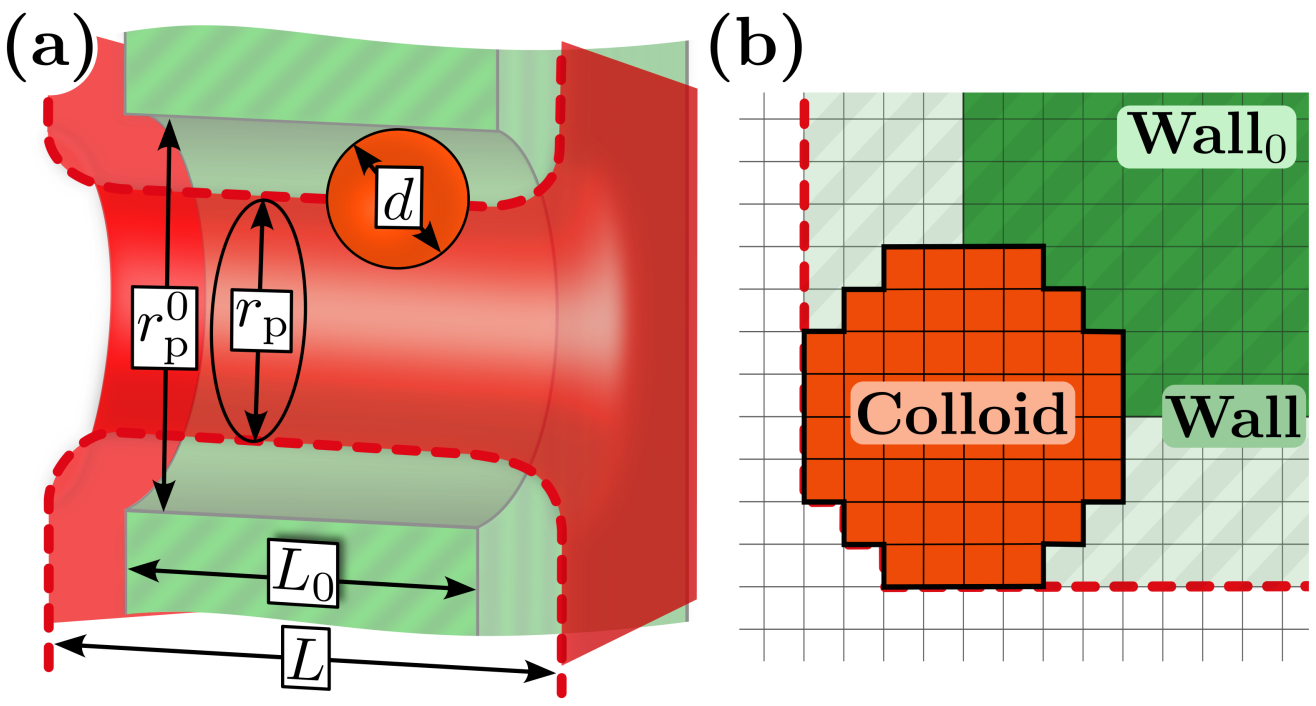

**Figure S10**: **Defining the impermeable wall on the cylindrical lattice. (a)** Effective pore size in continuous space. Perspective view of the real (green) and effective (red) pore shapes. In the cut through, the surface of the effective pore is traced with a dashed red line, and a colloid of size $d$ is shown in orange. **(b)** Effective pore size in discretized space. $rz$ plane view of the real (dark green) and effective (light green, delimited with a red dashed line) membrane wall shapes. The coarse-grained colloid is shown in orange.

The binary mask **Source**$[i, j]$ identifies the grid nodes at the boundary of the domain that act as the colloid source (see Figure S9a). These nodes lie outside the oblate ellipsoid defined by the above condition. The set is constructed as:

$$\textbf{Source}[i, j] = \begin{cases} 1, & \begin{aligned} &\text{if } \left\|(r_i, z_k) - (r_\mathrm{p}, -L/2)\right\| \\ &\ + \left\|(r_i, z_k) - (-r_\mathrm{p}, -L/2)\right\| \geq 2(r_\mathrm{max} - \delta r) \end{aligned} \\ 0, & \text{otherwise} \end{cases} \tag{S63}$$

This boundary condition ensures that the solution approximates that of an idealized system with a source located infinitely far away ($z \to -\infty$). We effectively reproduce the shape of iso-concentration lines in the exterior region of a bare pore, thereby maintaining a realistic flux geometry.

**Defining the colloid sink** Exploiting the system's symmetry, we impose a Dirichlet boundary condition $\psi = \psi_\mathrm{sink}$ at the $z+$ face of all finite volume elements with $k = N_z - 1$. This corresponds to placing the absorbing boundary condition at the mid-plane $z = 0$. Due to symmetry, there is no radial component of the flux at $z = 0$, so no flux crosses the $r-$ or $r+$ faces of these finite volume elements.

We define the set of finite volume elements that act as sinks as:

$$\textbf{Sink}[i,k] = \begin{cases} 1, & \text{if } k = N_z - 1, \\ 0, & \text{otherwise.} \end{cases} \tag{S64}$$

For these elements, the divergence term simplifies to account for the fixed potential at the $z+$ face and no flux at the radial faces:

$$\nabla_{i,k_0}\psi = \lambda^z D_{i,k_0}(\psi_{i,k_0} - \psi_{\text{sink}}) + 2\lambda^z D^{z-}_{i,k_0}(\psi_{i,k_0} - \psi_{i,k_0-1}) \tag{S65}$$

**Constructing the Laplace operator matrix.** The matrix **L** is constructed using a five-point stencil (Figure S9a) that adapts near boundaries. The diagonal entries are defined as

$$\boldsymbol{L}_{m,m} = \begin{cases} 0 & \text{if } m \in \textbf{Wall}, \\ 1 & \text{if } m \in \textbf{Source}, \\ -2\lambda^z D_m & \text{if } m \in \textbf{Sink}, \\ -\sum\limits_{m' \in \mathcal{N}_m} \boldsymbol{L}_{m,m'} & \text{otherwise,} \end{cases} \tag{S66}$$

where $\mathcal{N}_m$ is the set of valid neighbor indices:

$$\mathcal{N}_{i,k} = (\{(i \pm 1, k),\ (i, k \pm 1)\}) \cap \Omega, \tag{S67}$$

and node indices are flattened as $m = iN_z + k$ and $m' = i'N_z + k'$.

The off-diagonal entries $\boldsymbol{L}_{m,m'}$ represent finite-volume approximations of the Laplace operator (Eq. (S54)) taking account of the boundary conditions:

$$\boldsymbol{L}_{m,m+1} = \begin{cases} 0 & \text{if } m+1 \in \textbf{Wall} \\ D^{z+}\lambda^z & \text{otherwise} \end{cases} \tag{S68}$$

$$\boldsymbol{L}_{m,m-1} = \begin{cases} 0 & \text{if } m-1 \in \textbf{Wall} \\ 2D^{z-}\lambda^z & \text{if } m \in \textbf{Sink} \\ D^{z-}\lambda^z & \text{otherwise} \end{cases} \tag{S69}$$

$$\boldsymbol{L}_{m,m\pm N_z} = \begin{cases} 0 & \text{if } m \pm N_z \in \textbf{Wall} \text{ or } m \in \textbf{Sink} \\ D^{r\pm}\lambda^{r\pm} & \text{otherwise} \end{cases} \tag{S70}$$

The right-hand side vector $\boldsymbol{b}$ enforces the boundary condition from the source and the sink:

$$\boldsymbol{b}_m = \begin{cases} \psi_{\text{source}} & \text{if } m \in \textbf{Source} \\ \psi_{\text{sink}} & \text{if } m \in \textbf{Sink} \\ 0 & \text{otherwise} \end{cases} \tag{S71}$$

The resulting sparse linear system in Eq. (S58) is solved using the `scipy.sparse.linalg` package.

**Extracting pore resistance.** The total colloid flux was computed by summing the fluxes through the $z$+ faces of the finite volume elements at $k = N_z - 1$, corresponding to the pore cross-section at the mid-plane $z = 0$. Due to symmetry, the radial flux vanishes at this plane, so only the axial ($z$-direction) components contribute:

$$J_{\text{num}} = 2\pi \sum_{i=0}^{N_r-1} D_{i,k'} \frac{2(\psi_{i,k'} - \psi_{\text{sink}})}{\delta z}(2i+1)\delta r^2, \tag{S72}$$

where $k' = N_z - 1$ is the axial index of the mid-plane.

To determine the total pore resistance, we prescribe boundary conditions $\psi_{\text{source}} = 1.0$ and $\psi_{\text{sink}} = 0.5$. This corresponds to a concentration difference of $\Delta c = 1.0$ across the full system, assuming symmetry about the mid-plane:

$$R_{\text{num}} = \frac{\Delta c}{J_{\text{num}}} + 2R_{(|z_{\min} - L/2|,\, -\infty)} \tag{S73}$$

Here, $R_{(|z_{\min}-L/2|,-\infty)}$ is the analytically estimated resistance of the truncated semi-infinite reservoir, as given by Eq. (S20).

## Supplementary Note 7. Mapping our model to experimental data for biocolloid transport across NPCs

Figure S11 and Table S4 summarize the most relevant aspects of the published experimental quantifications of biocolloid transport rates across NPCs in the nuclear envelope of cells, along with partition coefficients for sticky colloids in phase-separated droplets of pure FG domain. The nucleus is represented as a well-mixed compartment of volume $V_{\text{nucleus}}$ bounded by an impermeable envelope that is perforated by $N_{\text{NPC}}$ nuclear pore complexes. The NPCs are assumed to be sufficiently far apart for transport through each NPC to be independent. Assuming (somewhat simplistically) that the nuclei are spherical, one can estimate a root-mean-square distances between NPCs of 0.44 $\mu$m for HeLa and 0.29 $\mu$m for yeast cells, based on the nuclear characteristics (Table S4). These values exceed the NPC transport channel diameter by at least 5-fold, and should indeed be sufficent for quasi-independent transport in most cases (as illustrated in Figure 6) (*52*). In mapping our model to the experimental data, we naturally ignore the chemical heterogeneity of FG domains and biocolloids. Instead, we demonstrate that this level of detail is dispensible for a basic yet quantitative description of the NPC permselectivity barrier.

**Extracting NPC resistances from the experimental data**

Table S4 highlights that the experimental conditions vary, and so do the reported transport rate observables. To enable direct comparisons, we first converted all observables into resistances of individual NPCs. To provide intuitively accessible numbers, we further considered translocation rates per NPC at a set probe colloid concentration of $\Delta c = 1\mu$M. The results are given in Tables S1 and S2, and Figure 8.

**Permeabilized human cells.** In several studies (*25–27*), the plasma membrane of human HeLa cells was digitonin-permeabilized whilst leaving the nuclear membrane intact, fluorescent probe colloids were then delivered and the initial rate of probe concentration increase in the nucleus was quantified. Here, the external solution represents a quasi-infinite reservoir with constant probe concentration $c_0$. Under these conditions, the probe concentration in the nucleus initially increases linearly with time, $c_{\text{nucleus}} \approx ktc_0$ for $t \ll k^{-1}$, with the measured rate constant $k$. From the rate constant, the experimentally determined resistance of each NPC in the nuclear envelope is obtained through:

$$R_{\text{exp}} = \frac{N_{\text{NPC}}}{kV_{\text{nucleus}}}. \tag{S74}$$

**Intact yeast cells.** Two studies (*28*, *29*) photobleached the nuclear pool of fluorescent probe colloids in intact *S. cerevisiae* cells, and then quantified the rate of influx of fluorescent probe molecules from the cytosol. Here, the nuclear and cytosolic volumes are both finite, and the rate

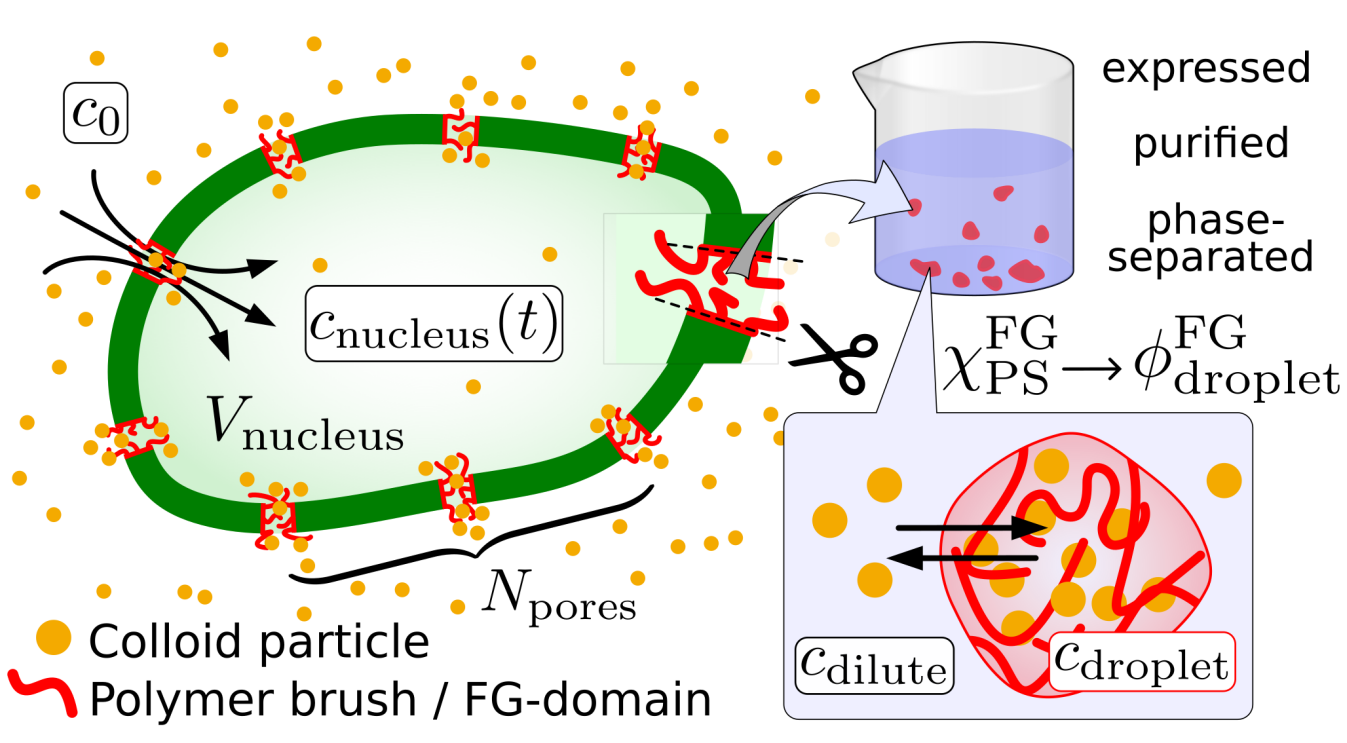

**Figure S11**: **Reductionist view of relevant pore-mediated equilibration experiments from Refs. (*25–29*). Left:** The nucleus with finite volume $V_{\text{nucleus}}$ is separated from a finite cytosol ($V_{\text{cytosol}}$) or a quasi-infinite bulk solution by an impermeable envelope (green contour) perforated by $N_{\text{NPC}}$ FG-domain filled NPCs. Biocolloids (yellow circles) are mobile, and a concentration difference $\Delta c$ drives their diffusive flux into the nucleus where they accumulate over time $c_{\text{nucleus}}(t)$. **Top right:** Isolated FG domain polymers phase-separate in a poor solvent with solvent strength $\chi_{\text{PS}}^{\text{FG}}$, forming droplets (red spheres) with polymer volume fraction $\phi_{\text{droplet}}$. **Bottom right:** Colloids equilibrate between the dilute and condensed droplet phases, characterized by the partition coefficient $P = c_{\text{droplet}}/c_{\text{dilute}}$.

constant defining the transport process is given by

$$k = t^{-1} \ln \left( \frac{1 + \frac{c_{\text{nucleus}}(t)}{c_{\text{cytosol}}(t)} \frac{V_{\text{nucleus}}}{V_{\text{cytosol}}}}{1 - \frac{c_{\text{nucleus}}(t)}{c_{\text{cytosol}}(t)}} \right) \tag{S75}$$

From the rate constant $k$, or equivalently the characteristic time $\tau = k^{-1}$, the resistance per NPC is obtained through:

$$R_{\text{exp}} = \frac{N_{\text{NPC}}}{k} \left( \frac{1}{V_{\text{nucleus}}} + \frac{1}{V_{\text{cytosol}}} \right) \tag{S76}$$

**Determining translocation rates per NPC.** From the expirimentally determined NPC resistances, the translocation rates per NPC for a given molar probe concentration are readily obtained as:

$$J_{\text{exp}} = \Delta c N_{\text{A}} / R_{\text{exp}} \tag{S77}$$

Naturally, this relation was also used to convert the theoretically predicted pore resistances into translocation rates. Here, the solvent viscosity and temperature (which affect $D_0$) were set to $\eta_{\text{S}} = 1.45 \times 10^{-3}$ Pa s and $T = 293$ K, respectively, matching the conditions used for the experiments with HeLa cells (*26*).

**Mapping model and experimental parameters for the transport of sticky colloids.**

Frey et al. (*25*) used probe protein variants (GFP and mCherry) with varying FG domain attraction but similar size, to correlate their translocation rates across HeLa cell NPCs with their partition coefficients $P$ in phase-separated droplets of pure FG domains of either Nup116 from *S. cerevisiae* or Nup98A from *T. thermophilia*.

To compare these data with the predictions of our model, we estimated the effective solvent quality $\chi_{\text{PS}}^{\text{FG}}$ of the pure FG domain droplets, and the effective interaction strength $\chi_{\text{PC}}^{\text{FG}}$ between the colloids and the pure FG domains, as follows.

**Estimating $\chi_{\text{PS}}^{\text{FG}}$.** Nup98A and Nup116 FG domains undergo spontaneous phase separation due to chain cohesiveness, demixing into a dilute FG-poor phase and condensed FG-rich droplets. The polymer concentration in the dilute phase, $\phi_{\text{FG dilute}}$, approximately equals the critical concentration for phase separation, and both chemical potentials and osmotic pressures are equal in the dilute and condensed phases (*24*, *68*).

The estimated critical concentration of Nup116 and Nup98A FG domains is $1\mu$g/ml, corresponding to a FG domain volume fraction in the dilute phase of $\phi_{\text{FG dilute}} \approx 10^{-6}$ (*21*). This extremely low value implies that the osmotic pressure is effectively zero, and allows to estimate the solvent quality from the condition for vanishing osmotic pressure (Eq. (16)):

$$\chi_{\text{PS}}^{\text{FG}} = \frac{-\ln(1-\phi_{\text{FG droplet}}) - \phi_{\text{FG droplet}}}{\phi_{\text{FG droplet}}^2} \tag{S78}$$

The concentration of Nup98A in phase-separated droplets was in a later study (*60*) quantifed at 474 ± 32 mg/mL, and corresponds to a volume fraction $\phi_{\text{Nup98A droplet}}$ between 0.31 and 0.51 considering the range of plausible densities (1.0 g/mL $< \rho <$ 1.4 g/mL, taking into account that some solvent may contribute to the effective FG domain volume). Similarly, an estimated concentration of Nup116 of 350 mg/mL (*21*) corresponds to $\phi_{\text{Nup116 droplet}}$ values between 0.23 and 0.38.

For simplicity, we used $\chi_{\text{PS}}^{\text{FG}} = 0.6$ to represent the Nup98A and Nup116 droplets, consistent with our assumption for all NPCs. This corresponds to a volume fraction of $\phi_{\text{FG droplet}} = 0.25$, i.e., close to the real values for Nup98A and Nup116. Within the range $0.2 < \phi_{\text{FG droplet}} < 0.5$, or $0.58 < \chi_{\text{PS}}^{\text{FG}} < 0.77$, the dependence of the transport rate on the partition coefficient $P$ is only moderately sensitive to the exact value of $\phi_{\text{FG droplet}}$ (Figure S12).

**Estimating $\chi_{\text{PC}}^{\text{FG}}$ for each probe protein.** The effective volume of the probe proteins was determined from the molecular mass $M_w$ as $V_{\text{probe}} = M_w/(N_{\text{A}}\rho)$, with Avogadro's number $N_{\text{A}}$ and 1.0 g/mL $< \rho <$ 1.4 g/mL. The effective size of the probes was taken as the diameter of a sphere of equivalent volume:

$$d_{\text{probe}} = \left(\frac{6}{\pi}V_{\text{probe}}\right)^{1/3} = \left(\frac{6}{\pi}\frac{M_w}{N_{\text{A}}\rho}\right)^{1/3} \tag{S79}$$

The molecular mass of probe proteins in this dataset was kept constant at ~ 28 kDa (Table S2), corresponding to an effective probe diameter between 4.0 and 4.5 nm, or between 5.2 and 5.9 when normalized by the segment length $a = 0.76$ nm. To simulate this system, we used $d_{\text{probe}} = 6$, i.e., the nearest even integer value compatible with our discretized model.

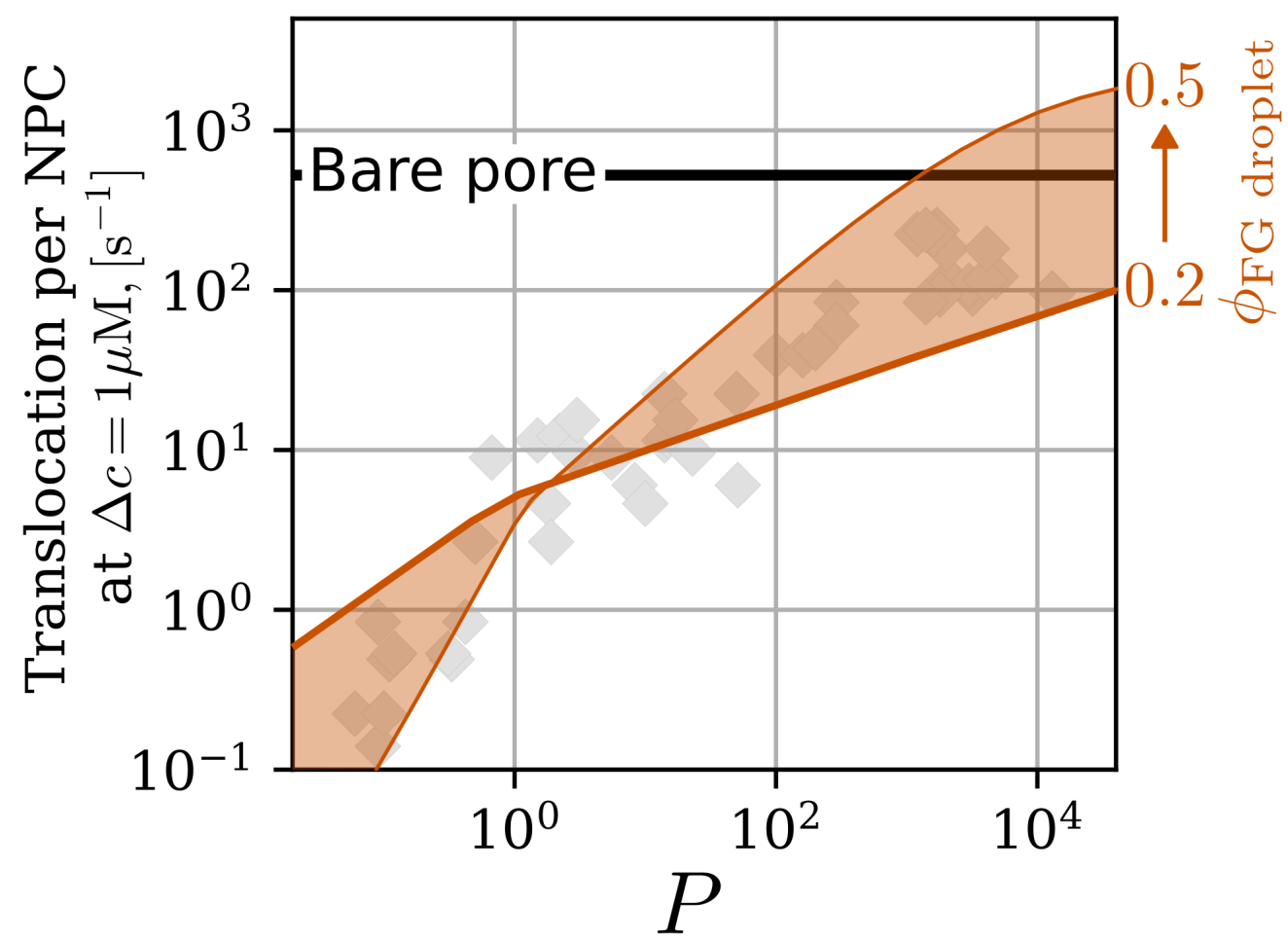

**Figure S12**: **Effect of polymer volume fraction in the FG-domain droplet on translocation rate - partitioning predictions.** Effect of the polymer volume fraction in the FG domain droplet, $\phi_{\text{FG droplet}}$, on the relationship between the translocation rate across NPCs and the partitioning into FG domain droplets. Theoretical predictions are presented as in Figure 8b for a colloid surface with a single sticky patch (orange lines and shaded area). All parameters are as in Figure 8b, except for $\phi_{\text{FG droplet}}$ which was varied between 0.2 and 0.5 (as indicated), corresponding to $\chi_{\text{PS}}^{\text{FG}}$ between 0.58 and 0.77. $\chi_{\text{PC}}^{\text{FG}}$ values (not shown) were mapped through Eqs. (S80-S81) to compute the translocation rates. The experimental data from Figure 8b are also shown (grey symbols), for comparison.

Since the osmotic pressure in the pure FG droplets is negligible, only surface interactions contribute to the insertion free energy $\Delta F_{\text{FG droplet}}$, such that:

$$\Delta F_{\text{FG droplet}} = \frac{\pi d_{\text{probe}}^2}{6} \cdot \gamma \left( \phi_{\text{FG droplet}}, \chi_{\text{PC}}^{\text{FG}}, \chi_{\text{PS}}^{\text{FG}} \right) \tag{S80}$$

$$P = \frac{c_{\text{droplet}}}{c_{\text{dilute}}} = e^{-\Delta F_{\text{FG droplet}}} \tag{S81}$$

This set of equations enabled mapping of $\chi_{\text{PC}}^{\text{FG}}$ values to the measured partition coefficients $P$ between the droplet and dilute phases for each probe protein (Figure 8b), with the established estimates of $\phi_{\text{FG droplet}} = 0.3$, $\chi_{\text{PS}}^{\text{FG}} = 0.6$, and $d_{\text{probe}} = 6$.

## Supplementary Tables

**Table S1**: **Translocation rates of non-sticky peptides and globular proteins across NPCs.** Based on experimental data reported in the literature. See references for details on the probe colloid constructs.

| Probe colloid | $M_w$ [kDa] | Translocation rate per NPC at $\Delta c$ = 1$\mu$M [$s^{-1}$] | Reference |
|---|---|---|---|
| Fluorescein-cysteine | 0.5 | 231 | *(27)* [a] |
| 11 amino acid peptide | 1.4 | 130 | |
| Insulin | 5.8 | 59 | |
| Aprotinin | 6.5 | 21.1 | |
| Ubiquitin | 8.5 | 8.7 | |
| Protein A z-domain | 8.2 | 9.9 | |
| Thioredoxin | 13.9 | 5.0 | |
| Lactalbumin | 14.2 | 3.54 | |
| Green fluorscent protein (GFP) | 27.0 | 0.50 [e] | |
| Phosphate binding potein (PBP) | 37.0 | 0.064 | |
| Maltose binding protein (MBP) | 43.0 | 0.054 [e] | |
| GFP | 26.8 | 1.11 [e] | *(29)* [b] |
| GFP-Protein A domain (PrA) | 34.2 | 0.268 | |
| GFP-2×PrA | 40.7 | 0.146 | |
| GFP-3×PrA | 46.8 | 0.092 | |
| GFP-4×PrA | 53.6 | 0.067 | |
| GFP-6×PrA | 66.8 | 0.040 | |
| GFP-Protein G C2 domain | 34.7 | 0.83 | |
| GFP-Protein G C2 and C3 domains | 42.3 | 0.25 | |
| MBP-GFP-MBP | 109 | 0.0091 | *(28)* [b] |
| MBP-GFP-2×MBP | 149 | 0.00250 | |
| MBP-GFP-4×MBP | 230 | 0.00116 | |
| MBP-3×GFP | 122 | 0.0052 | |
| MBP-4×GFP | 150 | 0.00368 | |
| MBP-5×GFP | 177 | 0.00197 | |
| Bovine serum albumin (BSA) | 68 | < 0.1 [c] | *(26)* [d] |
| GFP | 29.0 | 2 [e] | |
| mCherry | 28.0 | 0.140 | *(25)* [d] |
| GFP | 28.0 | 0.49 [e] | |
| efGFP_8R | 30.0 | 2.66 | |
| sffrGFP4 with 18×R→K mutations | 28.0 | 0.53 | |
| sffrGFP4 with 25×R→K mutations | 27.0 | 0.224 | |
| MBP | 43.0 | 0.0112 [e] | |
| MBP with K→R mutations | 43.0 | 0.45 | |

[a] Data for all probes in this source were included, except the profilin construct for which the molecular mass is unclear: the Stokes radius stated in the paper appears to small for the full human profilin 1 protein (15 kDa), suggesting only a part of the protein may have been used. [b] Data for all probes in this source were included. [c] A value of 0.1 $s^{-1}$, representing the upper bound, is shown in Figure 8a. [d] Selected data from this source were included, representing inert probes. [e] GFP and MBP were used in three and two studies, respectively, and all reported values were included in Figure 8a; the moderate variations in translocation rates for each protein are interpreted as experimental variations/uncertainties.

**Table S2**: **Translocation rates and partitioning of probe proteins in FG-domain droplets and NPCs.** Translocation rates across NPCs, and partition coefficients $P$ in phase-separated droplets of pure FG domains, of mCherry and variants of GFP with modified surface features, from Frey et al. (*25*). See original paper for details on the probe protein constructs.

| Probe protein | Protomer $M_w$ [kDa] | Translocation rate per NPC at $\Delta c = 1\mu\text{M}$ $[\text{s}^{-1}]$ [a)] | $P$ in Nup98A droplets [b)] | $P$ in Nup116 droplets [b)] |
|---|---|---|---|---|
| mCherry | 28 | 0.140 | - | 0.09 |
| GFP | 28 | 0.49 | 0.11 | 0.33 |
| efGFP_0W | 28 | 0.84 | 0.09 | 0.42 |
| efGFP_3W | 27.5 | 11.5 | 1.50 | 14 |
| efGFP_5W | 28 | 12.2 | 2.20 | 15 |
| efGFP_8F | 28.3 | 6.0 | 8.30 | 51 |
| efGFP_8L | 26.8 | 4.6 | 1.80 | 10 |
| efGFP_8I | 28.2 | 9.5 | 2.90 | 23 |
| efGFP_8M | 28.2 | 15.4 | 3 | 17 |
| efGFP_8R | 30 | 2.66 | 0.50 | 1.90 |
| sffrGFP4 | 29 | 22.4 | 14 | 50 |
| sffrGFP4 | 29 | 22.4 | 14 | 50 |
| sffrGFP4 18xR→K | 28 | 0.53 | 0.12 | 0.31 |
| sffrGFP4 25xR→K | 27 | 0.22 | 0.06 | 0.10 |
| sffrGFP4 | 29 | 22.4 | 14 | 50 |
| sffrGFP4 | 29 | 22.4 | 14 | 50 |
| sffrGFP5 | 28 | 9.0 | 0.67 | 5.50 |
| sffrGFP6 | 29 | 39.2 | 100 | 160 |
| sffrGFP7 | 29 | 43 | 200 | 200 |
| GFP_MaxR_5W | 28 | 116 | 2100 | 4000 |
| GFP_MaxR_8i | 27.6 | 182 | 2000 | 4100 |
| GFPNTR_2B7 | 27.1 | 224 | 1200 | 1600 |
| GFPNTR_7B3 | 26.3 | 238 | 1700 | 1400 |
| GFPNTR_3B1 | 28.5 | 60 | 290 | - |
| GFPNTR_3B7 | 27.5 | 106 | 3000 | - |
| GFPNTR_3B8 | 27.5 | 94 | 1800 | - |
| GFPNTR_3B9 | 28.3 | 122 | 4800 | - |

[a)] Translocation rates were calculated from the data in the original paper, as described in Supplementary Note 7.
[b)] Partition coefficients are reproduced from the original paper.

**Table S3**: **Analysis of the content of intrinsically disordered regions (IDRs) in FG nucleoporins in yest NPCs.**

| FG nucleoporin in *S. cerevisiae* [a] | Number of amino acids in IDR per protein [b] | Protein copy number per NPC [c] | Number of amino acids in IDRs per NPC |
|---|---|---|---|
| Nsp1 | 617 | 48 | 29616 |
| Nup49 | 251 | 32 | 8032 |
| Nup57 | 255 | 32 | 8160 |
| Nup145N | 433 | 16 | 6928 |
| Nup116 [d] | 960 | 16 | 15360 |
| Nup100 | 800 | 16 | 12800 |
| Nup60 [e] | 539 | 16 | 8624 |
| Nup1 | 857 | 16 | 13712 |
| Nup42 | 382 | 8 | 3056 |
| Nup159 | 685 | 16 | 10960 |
| Total | | 216 | 117248 |

[a] Nup2 was excluded as this protein is non-essential. [b] According to Yamada et al. (*66*). [c] According to Kim et al. (*18*). [d] The full C-terminal region up to amino acid 960 was considered IDR. [e] The full protein was considered disordered, based on prediction.

**Table S4**: **Transport-related quantities extracted from the experimental studies.**

| Study | Reported quantity | Nuclei from | $N_{\mathrm{NPC}}$ | $V_{\mathrm{nucleus}}$ [fL] | $V_{\mathrm{cytosol}}$ [fL] |
|---|---|---|---|---|---|
| Ribbeck et al. (*26*)<br>Mohr et al. (*27*)<br>Frey et al. (*25*) | Translocation rate per NPC at $\Delta c = 1\mu$M<br>Rate constant $k$<br>Rate constant $k$;<br>correlated with partitioning coefficient in phase-separated droplets of *S. cerevisiae* Nup116 and *T. thermophilia* Nup98A | HeLa cells | 2770 | 1130 | $\infty$ |
| Popken et al. (*28*)<br>Timney et al. (*29*) | Concentration ratio $c_{\mathrm{nucleus}}/c_{\mathrm{cytosol}}$ at $t$ = 1h<br>Characteristic time $\tau$ | *S. cerevisiae* | 161 | 4.8 | 60 |